\documentclass[useAMS,usenatbib]{mn2e}
\usepackage{graphicx}
\usepackage{amssymb}
\usepackage{subfigure}
\usepackage{MnSymbol}

\newcommand{\eg}{\hbox{e.g.,}}                 % e.g. NOT in italics
\newcommand{\degree}{\mbox{$^{\circ}$}}               % degrees
\newcommand{\idest}{i.e.}           % i.e. NOT in italics
\newcommand{\magnit}[2]{\mbox{$\mbox{\rm #1}^{\mbox{\rm\tiny m}}%
     \!\!\!.\!\,\, \mbox{\rm #2}$}}                   % magnitudes
\newcommand{\Msun}{\mbox{\,$M_{\odot}$\/}}          % solar mass
\newcommand{\oversim}[2]{\lower0.5ex\vbox{\baselineskip=0pt\lineskip=0.2ex
     \ialign{$\mathsurround=0pt #1\hfil##\hfil$\crcr#2\crcr\sim\crcr}}}
\newcommand{\density}{\mbox{ stars/pc$^{3}$}}           % km s-1

\usepackage{graphicx}
\usepackage{nicefrac}
\usepackage{lscape}
\usepackage{amsmath}
\title[Testing the universality of star formation]{Testing the universality of star formation - I. \\ Multiplicity in nearby star-forming regions}
\author[Robert R.~King et al.]{Robert R.~King$^{1}$\thanks{E-mail: rob@astro.ex.ac.uk}, Richard J.~Parker$^{2,3}$, Jenny Patience$^{1}$, and Simon P. Goodwin$^{3}$\\
$^{1}$Astrophysics Group, College of Engineering Mathematics and Physical Sciences, University of Exeter, Stocker Road, Exeter EX4 4QL, UK\\
$^{2}$Institute of Astronomy, ETH Z\"{u}rich, Wolfgang-Pauli-Strasse 27, 8093 Z\"{u}rich, Switzerland\\
$^{3}$Department of Physics and Astronomy, University of Sheffield, Hicks Building, Hounsfield Road, Sheffield, S3 7RH, UK}

\begin{document}

\date{Accepted ????. Received ????}

\pagerange{\pageref{firstpage}--\pageref{lastpage}} \pubyear{2010}

\maketitle

\label{firstpage}

\begin{abstract} 

We have collated multiplicity data for five clusters (Taurus, Chamaeleon~I, Ophiuchus,
IC348, and the Orion Nebula Cluster).  We have applied the same mass ratio (flux ratios of
$\Delta K\le$2.5) and primary mass cuts ($\sim$0.1--3.0\Msun) to each cluster and therefore
have directly comparable binary statistics for all five clusters in the separation range
62--620~au, and for Taurus, Chamaeleon~I, and Ophiuchus in the range 18--830~au.  We find
that the trend of decreasing binary fraction with cluster density is solely due to the high
binary fraction of Taurus, the other clusters show no obvious trend over a factor of nearly
20 in density.

With $N$-body simulations we attempt to find a set of initial conditions that are able to
reproduce the density, morphology and binary fractions of all five clusters.  Only an
initially clumpy (fractal) distribution with an initial total binary fraction of 73 per cent
(17 per cent in the range 62--620~au) is able to reproduce all of the observations (albeit
not very satisfactorily).  Therefore, if star formation is universal the initial conditions
must be clumpy and with a high (but not 100 per cent) binary fraction.  This could suggest
that most stars, including M-dwarfs, form in binaries.

\end{abstract}


\section{Introduction}

Star formation is one of the outstanding problems in astrophysics.  How stars form is
extremely interesting in itself, but also has huge implications for understanding galaxy
formation and evolution, and planet formation.

One of the major unsolved problems in star formation is the universality of the process: is
the difference between small, local star forming regions such as Taurus ($\sim 10^2
M_\odot$), and massive starburst clusters like 30 Doradus ($\sim 10^5 M_\odot$) merely one
of the level of star formation, or is there something fundamentally different between these
two extremes?

There is no evidence that the initial mass function (IMF) of stars varies systematically
between different environments \citep[see e.g.,][]{Luhman:2003,Bastian:2010}.  Such a
result is rather surprising and interesting, but it means that determinations of the IMF are
unable to probe the universality, or otherwise, of star formation. A more promising route
might be to search for differences between {\em primordial} binary populations -- if two
regions produce very different binary populations then this suggests that star formation was
different between these regions \citep[see][]{Duchene:2004,Goodwin:2010}.

Indeed, differences between the binary populations of different clusters have been observed.
Most famously, the binary fraction of Taurus \citep{Leinert:1993} is significantly higher
than the binary fraction of the Orion Nebula Cluster
\citep[hereafter ONC,][]{Prosser:1994,Petr:1998,Kohler:2006,Reipurth:2007} and approximately twice that seen
in the field \citep{Duquennoy:1991,Fischer:1992,Raghavan:2010}.

However, we know that dense regions will process their primordial binary populations and
what we see at later times may not reflect the primordial population. In a dense environment
encounters are common and binaries will tend to be destroyed \citep{Heggie:1975,Hills:1975}.
\citet{Kroupa:1995b,Kroupa:1995a} showed that it is possible to process a Taurus-like
primordial binary population into an Orion-like evolved population very quickly. However,
\citet*{Parker:2011c} suggest that even with dynamical processing the primordial binary
populations of Taurus and the ONC were probably different \citep[see
also][]{Kroupa:2011,Marks:2011}.

The problem is that it is difficult to `reverse engineer' the current binary population of a
cluster to determine the primordial population \citep[reverse population
synthesis,][]{Kroupa:1995a}.  A major aspect of this problem is that it is difficult to
compare the observed binary populations of different regions due to differences in the
separation range probed and the sensitivity to lower-mass companions between different
surveys.

In this paper we approach the problem of examining differences between primordial binary
populations with a two-fold approach.  Firstly, we construct (in as much as is possible) a
uniform comparison of binary fractions in the same separation ranges for five different
regions (Taurus, Oph/L1688, Cham~I, IC348, and the ONC).  Secondly, we attempt to simulate
their dynamical evolution and binary destruction as realistically as possible with both
smooth and clumpy initial conditions.

This paper is organised as follows.  In Section~2 we describe the observations of binarity
that we have used in our five different regions.  In Section~3 we construct as fair a
comparison as possible between the regions.  We summarise these results and discuss the
structure of each cluster in Section~4.  In Section~5 we introduce our $N$-body simulations
and compare these with the observational results.  In Section~6 we discuss our findings,
finally concluding in Section~7.

%%%%%%%%%%%%%%%%%%%%%%%%%%%%%%%%%%%%%%%%%%%%%%%%%%%%%%%%%%%%%%%%%%%%%%

\section{Cluster memberships and binarity}
\label{sect2}

We have chosen to compare binary surveys of young stars in five well-studied regions:
the Chamaeleon I cloud \citep{Lafreniere:2008}, Taurus \citep{Leinert:1993}, L1688 in
Ophuichus \citep{Ratzka:2005}, IC348 \citep{Duchene:1999} and the Orion Nebula Cluster
\citep{Reipurth:2007}.  The first four of these surveys all involved near-IR observations (3
$K$-band, 1 $H$-band) and so are most easily compared. The fifth survey (of the ONC) was
selected because it is the most comprehensive survey of this important, massive
star-formation region. These regions provide as broad a range in density as possible within
the confines of the Solar Neighbourhood which allows us to probe down to separations of a
few tens of au.

After summarising the most important past studies for each region, we identify the most
comprehensive binary studies with which we will make a comparison between the different
regions. For each region we report the multiplicity fraction (MF) and the corresponding
companion star fraction (CSF) defined as

\begin{equation}
MF = \frac{B + T + Q}{S + B + T + Q}
\end{equation}
\begin{equation}
CSF = \frac{B + 2T + 3Q}{S + B + T + Q}
\end{equation}

where $S$ is the number of single stars, $B$, $T$ and $Q$ are the numbers of binary, triple
and quadruple systems, respectively. Our $N$-body simulations do not produce systems with more than 
two components, so in this case the MF and CSF are equal to the binary fraction.

\begin{table}
  \centering
\renewcommand{\footnoterule}{}  % to avoid a line before footnotes
  \caption{A summary of the number of stars and densities calculated for each region.}
\label{tab:densities}
\begin{tabular}{lcccc}
  \hline
  \hline
  Region 	& \# of Stellar	&\multicolumn{3}{c}{Stellar Density (stars pc$^{-3}$)} 	\\
  		& Members	&  r$_{\nicefrac{1}{2}}$ & r$<$0.25\,pc	& r$<$0.10\,pc\\
  \hline
  Cham~I 	& 200		& 5.7$\pm$0.7	& 275$\pm$65    & 1190$\pm$530   \\
  Taurus 	& 215		& --- 		& 6.0$\pm$1.2   &       ---	 \\
  Oph 		& 295		& 236$\pm$27	& 610$\pm$180   & 1910$\pm$955   \\
  IC 348 	& 265		& 326$\pm$73 	& 1115$\pm$140  & 3820$\pm$1110  \\
  ONC 		& $\sim$1700	& 425$\pm$33	& 4700$\pm$290  & 22600$\pm$1200 \\
  \hline
\end{tabular}
{\it Note: }The density reported here for Taurus is calculated within a radius of 1\,pc from the
centre of L1495; the number of stellar members for the ONC is extrapolated from the number of COUP
sources within the half-number radius (r$_{\nicefrac{1}{2}}$) defined using the
\citet{Hillenbrand:1997} survey; and the number of stellar members for Taurus is for the Northern
filament only.
\end{table}

For each of the five star-forming regions we also report the most recent determinations of
the stellar membership.  With these data we form a rough estimate of the stellar densities. 
Note that to ensure consistency we exclude brown dwarfs from the density calculation and
focus on the more easily identified stellar population.

We measure the number of stars within the half-number, 0.1~pc and 0.25~pc radii from a
cluster `centre' determined from the average (`centre of mass') positions of all the stars.
We then assume that the third dimension is the same allowing us a basic estimate of the
stellar volume densities in each region as shown in Table~\ref{tab:densities}.  The
uncertainties on the densities are estimated by accounting for the Poisson errors on the
stars within each volume and the uncertainty on the distance to each region.   As we will
see later in Section~\ref{section:obs_morph}, several of these regions are far from
spherical and lack a proper `centre'.  However, we feel that these approximate densities
give a broad picture of the relative densities.

\subsection{Chamaeleon I}

\subsubsection{Membership}
 
% How have the known members been identified

To estimate the density of the young stellar cluster Cham~I, we have used the compilation of
known members presented by \citet{Luhman-SFHB:2008}. This member list was constructed from
the results of many past studies including surveys for H$\alpha$ emission, X-ray emission,
photometric variability and IR-excess emission. There have also been Cham~I members
identified using optical and near-IR imaging, due to the moderate optical extinction, and
the subsequent colour-magnitude diagram position of members relative to the contamination.

%Removal of known brown dwarfs

The known members include brown dwarfs with spectral types as late as M9.5. At the cluster
age of $\sim$2\,Myr \citep{Luhman:2004,Luhman:2007}, the sub-stellar limit occurs at an
approximate spectral type of M6. From the total of 237 known members of Cham~I, we are left
with 201 stellar members after removing those with spectral types later than M6 (as reported
by \citet{Luhman-SFHB:2008})

\begin{table*}
  \centering
  \caption{A summary of the separation ranges, contrasts and derived multiplicity fractions from
  each binary survey used.}
\label{tab:contrasts}
\begin{tabular}{llllcl}
  \hline
  \hline
  Region 	&  \multicolumn{2}{c}{Separation Range} &  Contrast &	Multiplicity Fraction & Reference  	\\
  
  		& (arcsec)	& (au)		& 	&  &\\
  \hline
  Taurus 	&  0.13--13.	&  18--1820	& $\Delta K \le2.5$	& 42$\pm$8\%	& \citet{Leinert:1993}	    \\
  Oph/L1688 	&  0.13--6.4	&  17--830      & $\Delta K \le2.5$	& 31$\pm$6\%	& \citet{Ratzka:2005}	    \\
  Cham~I		&  0.10--6.0	&  16--960	& $\Delta K \le3.1$	& 27$\pm$${5\atop4}$\%	& \citet{Lafreniere:2008}    \\
  IC 348 	&  0.10--8.0	&  32--2530     & $\Delta H \le6.5$	& 20$\pm$5\%	& \citet{Duchene:1999}	    \\
  ONC 		&  0.15--1.5	&  62--620	& $\Delta H \alpha \le5.0$	& 8.5$\pm$1.0\%	& \citet{Reipurth:2007}  \\
  \hline
\end{tabular}
\end{table*}
% Add footnote that this is all before normalisation
% Add note that contrast is not valid over full sep range. That is max. at large separation. 

% Talk about completeness

At an age of 2\,Myr, a 0.1$\Msun$ star is expected to have an apparent magnitude of
$K_S$$\simeq$11 at the distance of Cham~I. Given the 2MASS 10-$\sigma$ detection limit
($K_{S}\simeq$14.3), we would expect 2MASS to have detected all stars through A$_V$$<$30, much
larger than the estimated maximum extinction in Cham~I of A$_{V}=$5--10. Additionally, from
a very sensitive X-ray observation of the northern cluster of Cham~I, \citet{Feigelson:2004}
found no evidence for previously unreported members suggesting the known members in that
field are complete to 0.1$\Msun$. We therefore consider the membership list of
\citet{Luhman-SFHB:2008} to be essentially complete down to 0.1$\Msun$.

\subsubsection{Stellar density}

The Cham~I cluster comprises a northern and southern component with no obvious overall
centre.  For the northern component (centre $\alpha=167.47$\degree, $\delta=-76.515$\degree) we determined a
half-number radius of 0.85\degree\ or 2.37\,pc at a distance of 160\,pc \citep[see][for
discussion]{Luhman-SFHB:2008} and 0.59\degree\ or 1.65\,pc for the southern component
(centre 167.06\degree, -77.567\degree). This gives half-number radius densities of $\sim$2
and $\sim$5\density, respectively. Within radii of 0.25\,pc and 0.10\,pc the densities for
the southern component are 275$\pm$65\density\ and 1190$\pm$530\density\ (with little
difference for the northern component).

\subsubsection{Stellar binarity}

A number of studies have probed the binarity of this young cluster
\citep[e.g.,][]{Reipurth:1993,Ghez:1997}, but here we make use of the
\citet{Lafreniere:2008} study which acquired adaptive optics imaging of more than 50\% of
the known population. \citet{Lafreniere:2008} found 30 binary systems and 6 tertiary systems
in a sample of 126 Cham~I members with separations in the range 0.1--6.0\arcsec,
corresponding to 16--960\,au at a distance of 160\,pc. They report a multiplicity fraction
of 27$\pm$${5\atop4}$\% (CSF=32$\pm$${6\atop5}$\%) within this range, including only
companions where the flux ratio is above their 90\% completeness limits. Two apparent
companions were discounted due to a low probability of being bound to the primary and
follow-up spectroscopic observations which were inconsistent with the young age of the
region (\idest, they were likely distant background stars).

\subsection{IC\,348}

\subsubsection{Membership}

To estimate the stellar density in IC348, we have used the results of \citet{Luhman:2003}
who used optical and near-IR surveys, along with spectroscopic follow-up, to construct a
census which is complete well into the substellar regime. For consistency, we have
considered only those objects with spectral types of M6 or earlier, corresponding to sources
above the substellar limit in this $\sim$1--2\,Myr old cluster. After removing the brown
dwarfs, we are left with a stellar membership of 265 objects from the 288 known members.

\subsubsection{Stellar density}

Although subclustering is evident on spatial scales of $\sim$0.1\,pc \citep{Lada:1995},
IC348 shows a relatively symmetric radial profile. From a cluster centre of
$56.160$\degree,$+32.166$\degree, we have determined a half-number radius of 303\arcsec, or
$\sim$0.464\,pc at the cluster distance of 316$\pm$22\,pc \citep{Luhman:2003,Strom:1974},
which gives a stellar density in excess of 300\density - a determination which is
well-matched by those of \citet{Lada:1995} and \citet{Herbig:1998}.  Within radii of
0.25\,pc and 0.10\,pc, IC348 has stellar densities of 1115$\pm$138\density\ and
3819$\pm$1111\density, respectively.

\subsubsection{Stellar binarity}

In the first binary survey of IC\,348, \citet{Duchene:1999} reported the detection of
12 binary systems (and no higher order systems) from a sample of 66 targets systems using
the survey of \citet{Herbig:1998} to define cluster membership. They were sensitive to
binaries with separations down to 0.1\arcsec, or $\sim$32\,au at a distance of 316\,pc, and
their maximum separation of 8.0\arcsec was chosen to restrict the confusion between real
binary systems and background alignments. However, three apparent binaries were removed from
the sample as they were identified as likely background stars due to their large separations
and magnitude differences compared to the rest of the observed binary systems.

\citet{Duchene:1999} then use the known mass ratio distribution of the solar neighbourhood
from \citet{Duquennoy:1991} to estimate the number of undetected binary systems. They apply
this small correction to determine the likely number of total binaries in IC348 and so
derive a total multiplicity fraction of 19$\pm$5\% (CSF=19$\pm$5\%, since no $n>2$ systems
were found) within a separation range of 0.1--8.0\arcsec, or 32--2530\,au at a distance of
316\,pc.  After accounting for stars rejected as non-members, but which appear in the more
recent \citet{Luhman:2003} compilation, we determined that the \citet{Duchene:1999} study
detected 14 binaries from 71 targets systems, giving a multiplicity fraction of 20$\pm$5\%
over a separation range of 32--2530\,au.

\subsection{The ONC}
\label{sect:onc}

\subsubsection{Membership} 

Although a very well-studied region, there is no published list of confirmed stellar
members down to the substellar limit which covers more than the centre of this rich star
cluster. Therefore, to estimate the stellar density of the ONC, we
have used the complementary membership lists of \citet{Hillenbrand:1997} (hereafter H97) and
the $Chandra$ Orion Ultradeep Project \citep[COUP,][]{Getman:2005}. The H97 observations
cover a large area ($\sim$0.5$\times$0.5\degree), but do not probe down to the substellar
limit, while the COUP list is relatively complete to below $\sim$0.1\Msun, but covers only
the central $\sim$17\arcmin$\times$17\arcmin.

\subsubsection{Stellar density}

The ONC shows a slightly north-south elongated structure, but is centrally concentrated with
a dense core. We therefore used the H97 list to determine a half-number radius of
390\arcsec\ centred on 83.8185\degree, -5.3875\degree, corresponding to $\sim$0.78\,pc at a
distance of 414$\pm$7\,pc \citep{Menten:2007}.  From this we extrapolate that the ONC has a
stellar population of $\sim$1700 stars and within the half-number radius has a stellar
density of 425$\pm$33\density. This increases to 4700$\pm$290\density\ within 0.25\,pc of
the cluster centre and to 22,600$\pm$1200\density\ in the inner 0.1\,pc.

\subsubsection{Stellar binarity}

There have been several studies of the binarity of stars in this nearby massive star-forming
region. \citet{Prosser:1994} reported an estimated {\it binary fraction} of $\sim$11\% in
the range 0.1--1.0\arcsec\ (42--420\,au). \citet{Petr:1998} then used high angular
resolution near-IR imaging to probe ONC binaries and reported a {\it binary fraction} of
5.9$\pm$4.0\% in the separation range 0.14--0.5\arcsec (58--207\,au), but they were hampered
by very low numbers (only four binaries were detected).

To provide the most comprehensive sample for comparison with other regions, we have chosen
to use the more recent and wider-field HST survey of \citet{Reipurth:2007} which imaged over
1000 stars, of which 781 have a high membership probability. They found 78 multiple systems
with separations in the range 0.1--1.5\arcsec\ (42--620\,au) and from the density of stars
they estimated that 9 of their observed binaries were a result of projection effects.
\citet{Reipurth:2007} report a background-corrected multiplicity fraction of 8.5\%$\pm$1.0\%
(CSF=8.8\%$\pm$1.1\%) in the range 0.15--1.5\arcsec, or 62--620\,au\footnote{The separation
range quoted here is different to that given by \citet{Reipurth:2007} as we use the a newer
distance from \citet{Menten:2007}, supported by \citet{Jeffries:2007} and
\citet{Mayne:2008}.} This includes companions with flux ratios of up to
$\Delta$H$\alpha$$\sim$6\magnit

\subsection{Ophiuchus}

\subsubsection{Membership}

Due to the large and dispersed nature of the $\rho$ Ophiuchi complex we have chosen to focus
on the main cloud, L1688. A census of the known members of this core was presented by
\citet{Wilking:2008} which they believe to be `essentially complete' for class II and III
objects. This is supported by their comparison of the X-ray luminosity functions of L1688 to
that of the ONC from the deep COUP study of \citet{Feigelson:2005}.  For consistency with
the other regions studied here, we have used the \citet{Wilking:2008} list of candidate
brown dwarfs to remove those from the member list, leaving 295 known stellar members in
L1688.

\subsubsection{Stellar density}

While the stellar density of the Ophiuchus association taken as a whole is relatively low,
the density of the L1688 core is approximately an order of magnitude higher. The L1688 core
shows significant sub-clustering with no obvious overdensity at the centre.
\citet{Wilking:2008} summarise the various distance estimates for the $\rho$ Ophiuchus cloud
(120--145\,pc) and so, similarly, we adopt a distance of 130\,pc to L1688.  Using a cluster
centre of $246.727$\degree, $-24.44$\degree, we have determined a half-number radius of
0.234\degree, or $\sim$0.5\,pc at the distance of L1688, which results in an mean density of
236$\pm$27\density. Within radii of 0.25\,pc and 0.10\,pc, L1688 has stellar densities of
611$\pm$183\density\ and 1910$\pm$955\density, respectively.

\subsubsection{Stellar binarity}

In a lunar occultation and direct imaging search for binary stars, \citet{Simon:1995}
targeted 35 pre-main-sequence stars in Ophiuchus, but the small sample size frustrated their
attempts to compare with surveys of Taurus. More recently, \citet{Ratzka:2005}
presented a binary survey of 158 young stellar systems in the $\rho$ Ophiuchus
molecular clouds, centred on the dark cloud L1688. They reported a multiplicity fraction of
29.1$\pm$4.3\% within a separation range of 0.13--6.4$\arcsec$ (corresponding to 17--830\,au
at 130\,pc) where their observations were fully sensitive to flux ratios $\ge$0.1, but with
a significant number of companions with higher flux ratios.

However, if we consider only those systems which have been identified as members of Cham~I
in the recent \citet{Wilking:2008} census, then we are left with 106 systems, of which 32
are binaries and 3 tertiary systems.  \citet{Ratzka:2005} also analysed the contribution of
the background to the observed number of companions and determined that for their sample
there should be three unidentified non-bound systems. Similarly, for our reduced sample,
there should be two systems where the apparent companion is not bound, resulting in a
multiplicity fraction of 31$\pm$6\% (CSF=34$\pm$6\%) within a separation range of
17--830\,au.

\subsection{Taurus}

\subsubsection{Membership}

A census of the known stellar and substellar pre-main sequence members of the Taurus-Auriga
association was compiled by \cite{Kenyon:2008} and updated by \citet{Luhman:2009}. The
completeness of this sample was investigated by \citet{Luhman:2009} who reported that the
regions covered by the XEST survey \citep[][where complimentary optical and IR surveys
exist]{Gudel:2007} are complete for class I and II stars and complete down to 0.02\Msun\ for
class II brown dwarfs. Deep, wide-field, optical, near-IR
\citep{Briceno:2002,Luhman:2004,Guieu:2006} and Spitzer imaging surveys \citep{Luhman:2010}
of Taurus provide a high level of completeness into the substellar regime across the region.
Therefore, to provide an essentially complete {\it stellar} membership list we have removed
objects with spectral types later than M6, corresponding to sources
above the substellar limit in this $\sim$1--2\,Myr old cluster, leaving 292 stars.

\subsubsection{Stellar density}

Due to the dispersed nature of the young stars in the $\sim$1--2\,Myr Taurus-Auriga
association, it is not useful to define densities within a half number radius or within
radii as small as 0.25\,pc. We therefore report the average surface density of
$\sim$0.4$\pm$0.1 stars/pc$^{2}$ for the northern filament (defined here as
$62$\degree$<\alpha<72$\degree, $22$\degree$<\delta<31$\degree) and the volume density of
6.0$\pm$1.2\density\ within a radius of 1\,pc from the centre of the densest core (L1495,
centre $64.6$\degree,$+28.40$\degree) using a distance of 140$\pm$14\,pc
\citep{Wichmann:1998,Kohler:1998}.

\subsubsection{Stellar binarity}

Although a number of authors have reported binary statistics for young stars in Taurus
\citep{Ghez:1993,Duchene:2004}, in some cases probing down to below 1\,au
\citep{Simon:1992,Simon:1995}, here we make use of the binary survey of Taurus
presented by \citet{Leinert:1993} which surveyed over 100 stellar members. We do not use the
survey of \citet{Kohler:1998} as the weak-lined T Tauri stars identified though their X-ray
emission are more widespread across the region than the majority of the confirmed Taurus
members, suggesting that they may be a separate population. That said, \citet{Kohler:1998}
find no significant difference in binarity between the weak-lined and classical T Tauri
stars in the two samples. \citet{Leinert:1993} reported a multiplicity fraction of
42$\pm$6\% for their observations which were sensitive to systems with a flux ratio of up to
$\Delta$$K$=2.5 over a separation range of 0.13--13.0\arcsec, corresponding to 18--1820\,au.
The contribution of the background was examined and two apparent companions were discounted
as their large separations and colours identified them as background stars. No other
projected companions were expected in their sample.

If we consider only the targets within the area of the northern filament (as described
above), \citet{Leinert:1993} find 27 binary, 2 triple and 1 quadruple system from a total of
72 surveyed systems, giving a multiplicity fraction of 42$\pm$8\% (CSF=47$\pm$8\%) within a
separation range of 18--1820\,au.

\begin{figure}
    \resizebox{\hsize}{!}{\includegraphics{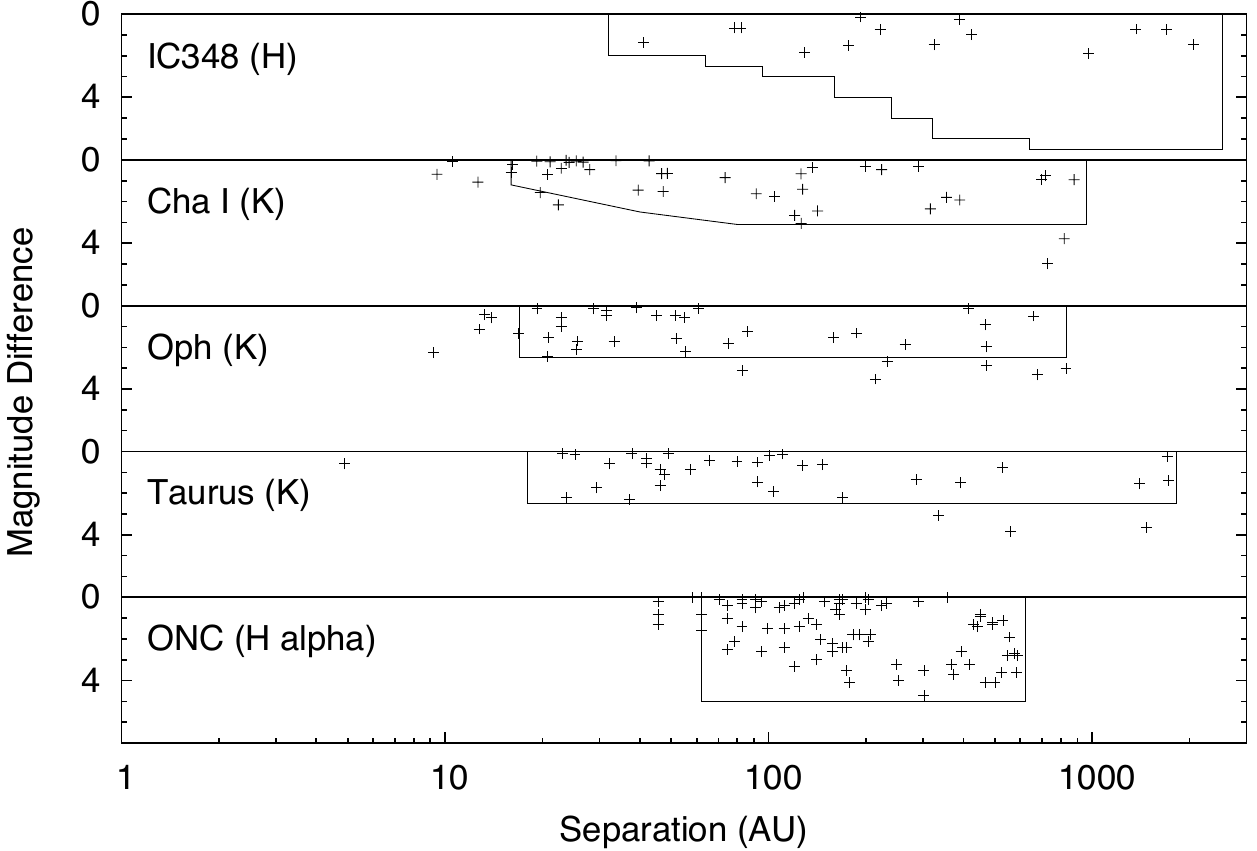}}
    \caption{The contrast of each multiple system found in the five surveys 
    shown as a function of separation. The filled lines demarcate the 
    completeness of each survey.  The labels identify the clusters and the 
    filter used in the observations.}
    \label{fig:contrasts}
\end{figure}

\section{Comparison of Observed Stellar Binarity}
\label{sect:comp}

% DISUSS RELATION OF DELTA k TO DELTA h AND DELTA Ha/R
\subsection{Contrast sensitivities}

To enable a fair comparison of the various binary surveys we must determine the
contrast ratio to which each survey was sensitive. Table\,\ref{tab:contrasts} lists the
maximum contrast ratio for each survey in the passband employed while
Fig.\,\ref{fig:contrasts} shows how these vary with physical projected distance from the primary star. For
simplicity, we aimed to use only surveys carried out in the $K$-band, but this was not
possible for IC\,348 and would have severely restricted the sample for the ONC, which used
an $H$-band and the NICMOS F658N filter respectively. For the surveys carried out in the
$K$-band, a common contrast cut of $\Delta K=2.5$ was chosen.

\begin{figure}
    \resizebox{\hsize}{!}{\includegraphics{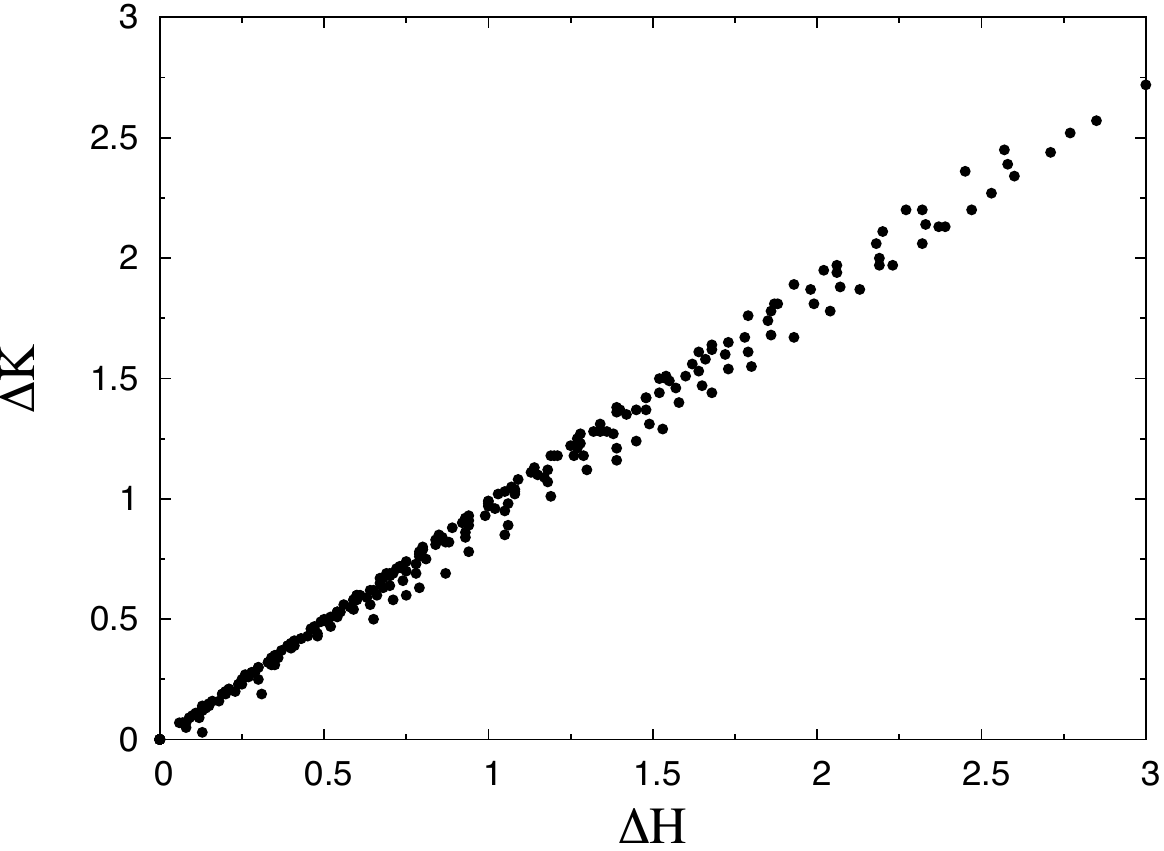}}
    \caption{The magnitude difference in the $H$ and $K$ bands between primary and secondary stars 
    using the predicted brightnesses from the 1\,Myr \citet{Siess:2000} evolutionary model for
    primary masses in the range 0.1--3.0\Msun\ and mass ratios of 0.1--1.0.}
    \label{fig:deltaH-deltaK}
\end{figure}

To determine a conversion of the contrast in the $H$ and F658N filters to the $K$-band, we
have used the theoretical models of \citet{Siess:2000}. By considering primary stars at
1\,Myr with masses in the range 0.1--3.0\Msun\ and mass ratios of 0.1--1, we were able to
predict the range of magnitude differences expected. Figure\,\ref{fig:deltaH-deltaK} shows
the relation between model magnitude differences in the $H$ and $K$-bands for this sample of
possible systems. This clear linear relation allows us to convert our chosen common $K$-band
contrast limit to an $H$-band contrast limit.

\begin{figure}
    \resizebox{\hsize}{!}{\includegraphics{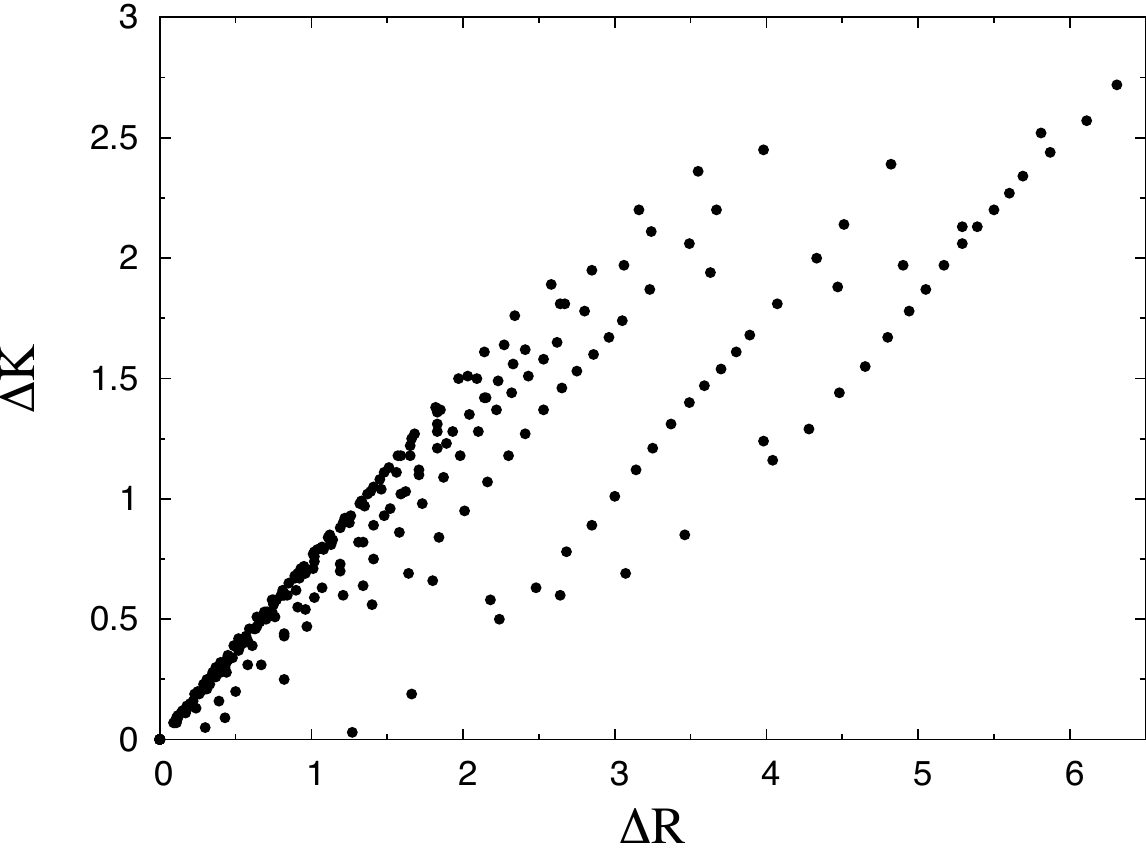}}
    \caption{The magnitude difference in the $R$ and $K$ bands between primary and secondary stars 
    using the predicted brightnesses from the 1\,Myr \citet{Siess:2000} evolutionary model for
    primary masses in the range 0.1--3.0\Msun\ and mass ratios of 0.1--1.0.}
    \label{fig:deltaR-deltaK}
\end{figure}

For the F658N filter, the situation is complicated by the lack of reported magnitudes in
that filter for the theoretical models. However, comparisons with the IPHAS H$\alpha$, $r'$
and Cousins R-band indicate that there is a near linear relation between magnitudes in these
bands. We therefore compare the magnitude differences in the $R$ and $K$ bands to determine
the appropriate contrast cut for the \citet{Reipurth:2007} survey of the ONC.
Figure\,\ref{fig:deltaR-deltaK} shows the relation between model magnitude differences in
the $R$ and $K$-bands for the primary masses and mass ratios described above. Although the
structure observed does not provide such a clear correlation as for the $H$ and $K$-bands,
our $\Delta$$K$ limit of 2.5 allows most of this structure to be ignored and gives a
contrast of F658N$\simeq$5. We also note that the majority of the \citet{Reipurth:2007}
binaries have contrasts of $\Delta$$R<3.0$. This then allows us to apply a consistent
contrast limit across the five surveys ($\Delta K=2.5$ $=>$ $\Delta H=2.7$,
$\Delta$F658N$=5$).

\subsection{Primary mass and mass ratio}

In the field the binarity of stars appears to decrease with primary mass
\citep{Duquennoy:1991,Fischer:1992,Lada:2006}. Simulations suggest that dynamical destruction
is relatively insensitive to primary mass (at least from M-dwarfs to G-dwarfs), and so this
mass-dependence may reflect a primordial mass-binarity relationship
\citep[see][]{Parker:2011b}.  Therefore, in addition to matching contrast ratios between
different surveys, we must ensure we cover the same masses of stars to avoid introducing a
possible bias in the observed binarity. To do this we have used the spectral type reported
in the various membership and binary survey papers for the survey targets to set upper and
lower mass limits
\citep{Hillenbrand:1997,Lafreniere:2008,Luhman:1999,Luhman:2003,Luhman:2009}. Using the
spectral type range of G5--M5.5 common to all five surveys, we have limited our comparison
to primary stars with masses of $\sim$0.1--3.0\Msun\ assuming an age of 1\,Myr
\citep{Siess:2000}.

\begin{figure}
    \resizebox{\hsize}{!}{\includegraphics{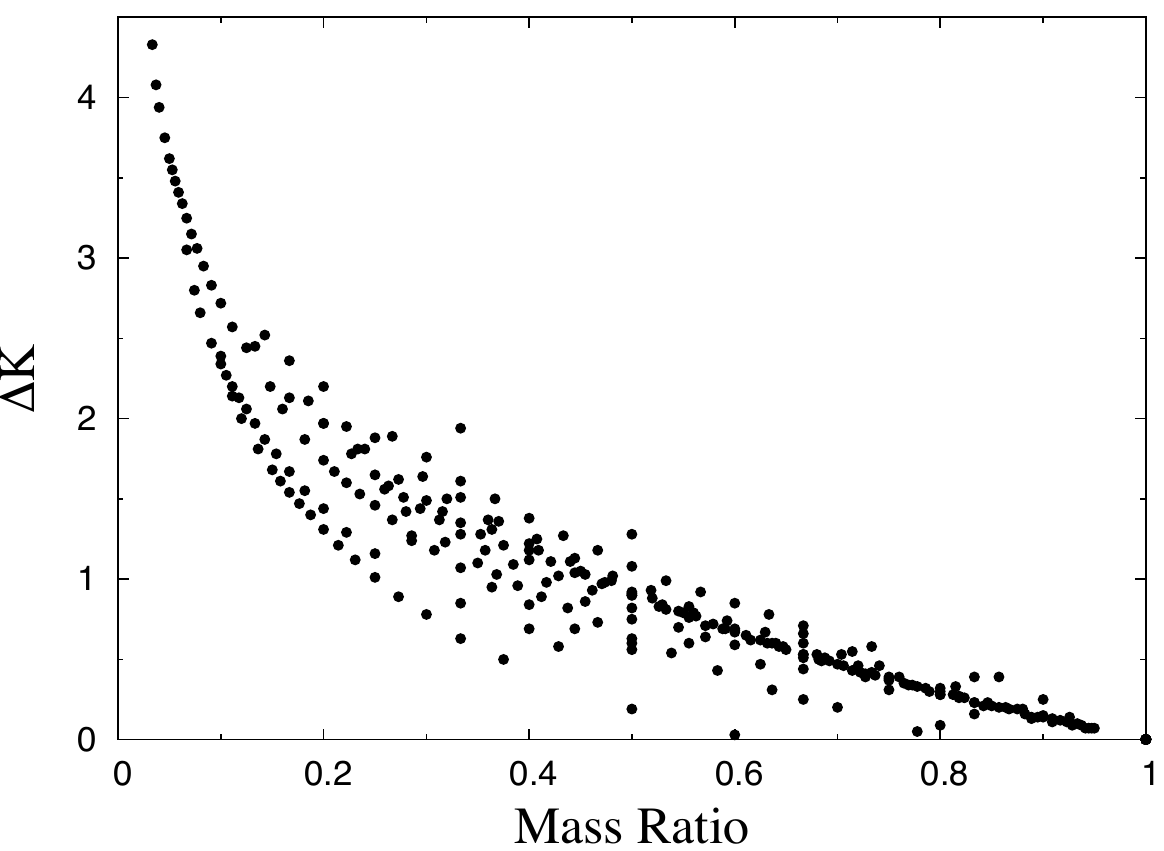}}
    \caption{The $K$-band magnitude difference as a function of mass ratio for systems with primary
     and secondary masses in the range 0.1--3.0\Msun\ from the 1\,Myr \citet{Siess:2000} 
     evolutionary model.}
    \label{fig:q-deltaK}
\end{figure}

While, for our survey comparison, we need not explicitly set limits on the range of mass
ratios probed, we can use the theoretical models of \citet{Siess:2000} to estimate the mass
ratios given the contrast limit of $\Delta K$=2.5. As shown in Fig.\,\ref{fig:q-deltaK}, the
contrast limit we have adopted is approximately equivalent to a mass ratio of 0.1 for
0.1--3.0\Msun\ primary stars at an age of $\sim$1\,Myr. We note however, that there will be
a small bias due to differing levels of completeness in the binary surveys (assuming a
variation with primary mass), \idest, one survey may have surveyed a larger fraction of
lower mass stars than another and so may find a slightly decreased binarity.

\subsection{Separation sensitivities}

The final cuts necessary to compare the binary surveys are to the separation ranges
probed. As the ONC is the farthest and densest of our regions, it sets a limit on the upper
and lower separation probed by all five surveys. However, applying this to all five surveys
would severely restrict the number of binary systems. We therefore present a comparison of
the five surveys with three different separation range cuts. Cut 1 (18--830\,au) allows us
to compare the widest possible separation range for Cham~I, Ophiuchus and Taurus; cut 2
(32--830\,au) also includes IC\,348; and cut 3 (62--620\,au) allows a comparison of all five
regions.

\section{Observational Results}

In Table~\ref{tab:results} we show the comparable multiplicity fractions for the five
regions in the three separation ranges.  For each region the companions and targets of the
binary surveys have been removed where the spectral types are not within the range G5--M5.5
and where the magnitude difference exceeds 2.5$^{\rm{m}}$. In the case of the ONC, the
spectral type information was not available so no mass cuts have been made to the sample. 
This is likely to mean the binarity is higher than it should be for this
comparison due to the (postulated) increasing binarity with stellar mass \citep[see,
\eg][]{Raghavan:2010,Parker:2011b}.

For comparison, a log-normal field G-dwarf-like distribution ($\mu_{{\rm log}~a} = 1.57$,
$\sigma_{{\rm log}~a} = 1.53$, with a 60 per cent {\em total} binary fraction) would
have a binary fraction of 24 per cent in the range $18 - 830$~au, 20 per cent in the
range $32 - 830$~au, and 14 per cent in the range $62 - 620$~au.  Similarly, for an
M-dwarf-like distribution with a {\em total} binary fraction of 40 per cent (and the same
log-normal parameters) the binary fractions would be 16, 13, and 9 per cent
respectively.  As most stars in our samples are M-dwarfs the most reasonable comparison is
to the M-dwarf-like field distribution.

In the $18 - 830$~au range all clusters are over-abundant in binaries compared to the
M-dwarf-like field distribution, and Taurus very significantly so.  In the $32 - 830$~au
range only Taurus and Cham~I are over-abundant, and in the $62 - 620$~au range only Taurus
has a significant excess.  We will return to this in the discussion.

\subsection{Binarity variations with stellar density}

\begin{figure}
    \resizebox{\hsize}{!}{\includegraphics{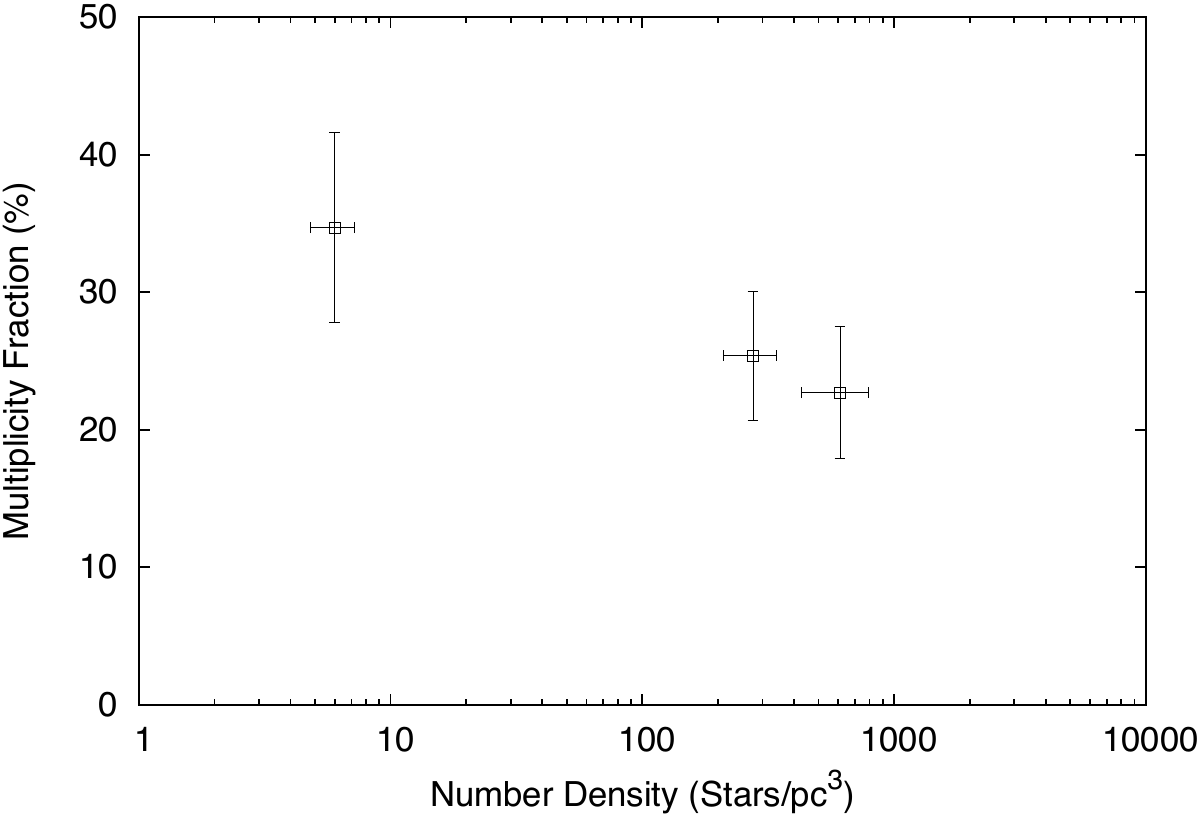}}
    \caption{The stellar multiplicity fractions of Taurus, Cham~I and Ophiuchus against stellar density when we
    consider the same contrast cuts, stellar masses and a separation range of 18--830\,au. The
    densities are calculated within a projected distance of 0.25\,pc from the cluster centre, except
    in the case of Taurus where a radius of 1\,pc is used. }
    \label{fig:binary_frac_vs_density_cut1}
\end{figure}

\begin{figure}
    \resizebox{\hsize}{!}{\includegraphics{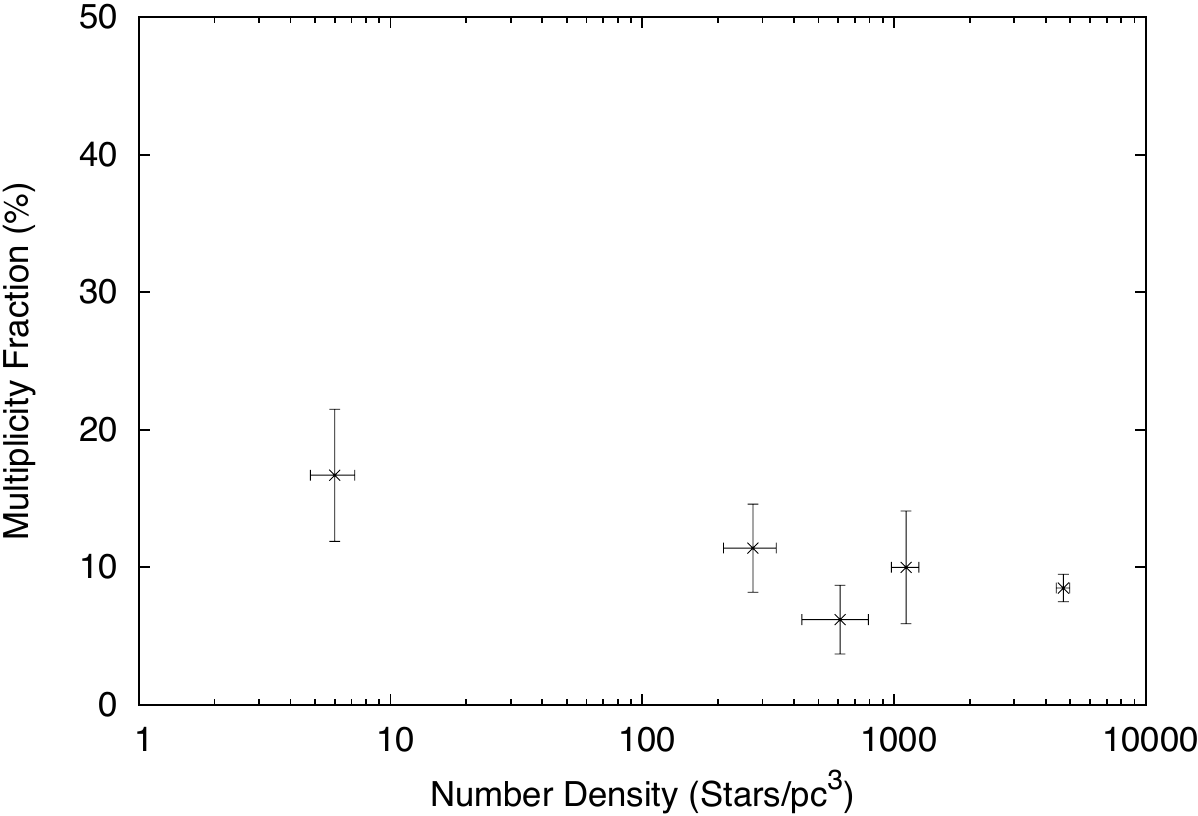}}
    \caption{The stellar multiplicity fractions of all five regions against stellar density when we
    consider the same contrast cuts, stellar masses and a separation range of 62--620\,au. The
    densities are calculated as for Fig. \ref{fig:binary_frac_vs_density_cut1}.}
    \label{fig:binary_frac_vs_density_cut3}
\end{figure}

\begin{table}
  \centering
  \caption{Stellar densities and multiplicity fractions for the three separation ranges.}
\label{tab:results}
\begin{tabular}{lcc}
  \hline
  \hline
  Region	       &  Stellar Density      &  Multiplicity fraction   \\
         	       &  (star pc$^{-3}$)     &  (per cent)   \\
  \hline
  \multicolumn{3}{c}{Separation = 18--830\,au} 			    \\
  \hline							
  Taurus	       	& 6.0 $\pm$ 1.2	       & 34.7 $\pm$ 6.9    \\
  Cham~I	       		& 275 $\pm$ 65	       & 25.4 $\pm$ 4.7    \\
  Ophiuchus/L1688      	& 610 $\pm$ 180	       & 22.7 $\pm$ 4.8    \\
  \hline
  \multicolumn{3}{c}{Separation = 32--830\,au} 			    \\ 
  \hline
  Taurus	       	& 6.0 $\pm$ 1.2	       & 29.2 $\pm$ 6.4    \\
  Cham~I	       		& 275 $\pm$ 65	       & 17.5 $\pm$ 3.9    \\
  Ophiuchus/L1688      	& 610 $\pm$ 180	       & 14.4 $\pm$ 3.9    \\
  IC 348	       	& 1115 $\pm$ 140	& 11.7 $\pm$ 4.4    \\
  \hline
  \multicolumn{3}{c}{Separation = 62--620\,au} 			    \\ 
  \hline
  Taurus	       	& 6.0 $\pm$ 1.2	       & 16.7 $\pm$ 4.8    \\
  Cham~I	       		& 275 $\pm$ 65	       & 11.4 $\pm$ 3.2    \\
  Ophiuchus/L1688      	& 610 $\pm$ 180	       & 6.2 $\pm$ 2.5    \\
  IC 348	       	& 1115 $\pm$ 140 	& 10.0 $\pm$ 4.1    \\
  ONC	       	       	& 4700 $\pm$ 290	& 8.5 $\pm$ 1.0    \\
  \hline
\end{tabular}
\end{table}

Given that binary fractions are thought to evolve due to the dynamical destruction of
binaries it is usually assumed that there should be a decrease in the binary fraction with
stellar density \citep{Kroupa:1995b,Kroupa:1995a,Parker:2009}.  This relationship is thought
to be seen in the significant differences in the binary fractions of Taurus and the ONC
($\sim 17$ per cent versus $\sim 8.5$ per cent respectively).

In Figs.~\ref{fig:binary_frac_vs_density_cut1} and~\ref{fig:binary_frac_vs_density_cut3} we
show the binary fractions against density for the range 18--820~au (for Taurus, Cham~I, and
Ophiuchus), and 62--620~au (for all five clusters).  In all cases the density is calculated
within a projected radius of 0.25~pc, except Taurus which is within a 1~pc projected radius.

There is a relationship between binary fraction and density, but this relationship is driven
almost entirely by Taurus.  In Fig.~\ref{fig:binary_frac_vs_density_cut3} in particular, the
relationship without Taurus is weak at best.  This is rather unexpected as the four clusters
(without Taurus) span more than a decade in density.

\subsection{Observed morphologies}
\label{section:obs_morph}

We have used the stellar density of clusters detailed above, but there are problems
associated with the determination (or meaning) of an average density in at least two of our
clusters (Taurus and Cham~I).  In smooth distributions taking a typical stellar density as a
measure of the proximity of stars and the likelihood of encounters is perfectly reasonable. 
However, at least two of the clusters we are examining are far from smooth and it is
questionable to what extent a `stellar density' is a meaningful concept.

\begin{figure}
    \resizebox{\hsize}{!}{\includegraphics{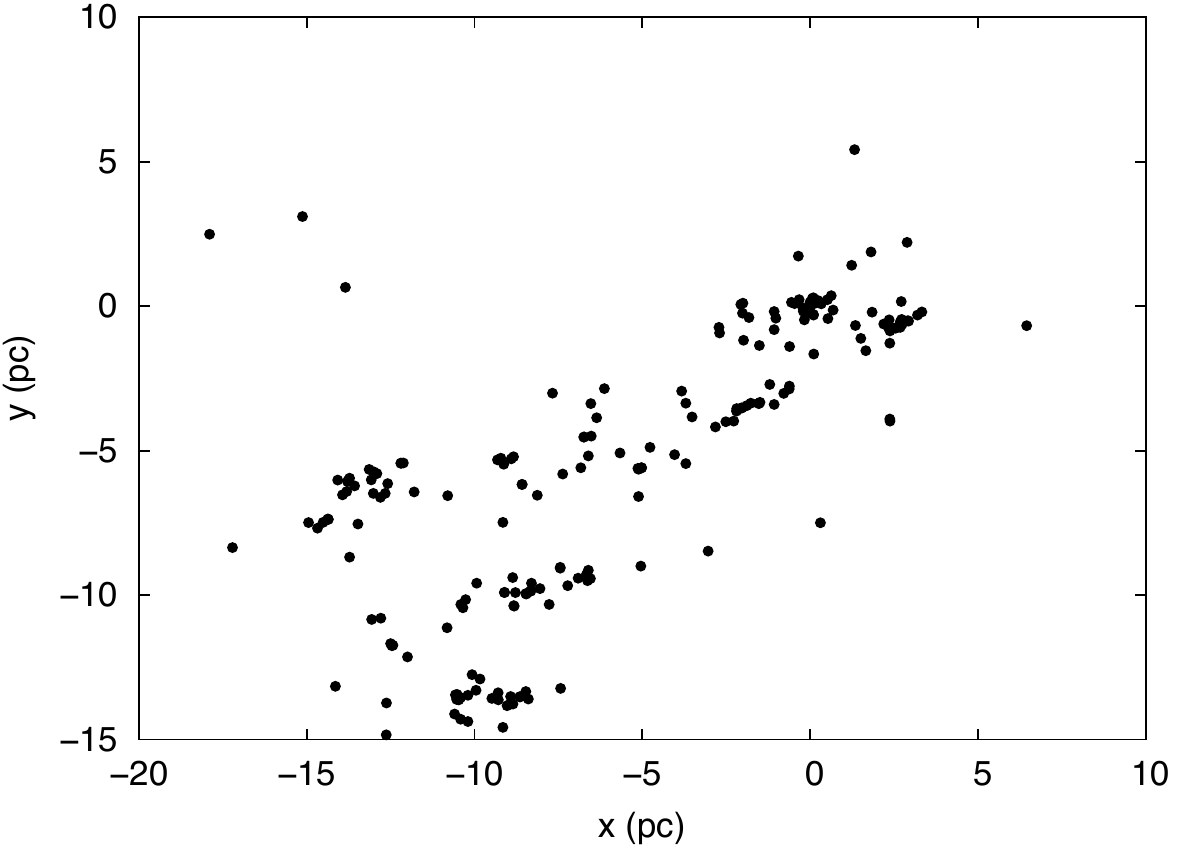}}
    \caption{The locations of each of the members of the northern filament of the
    Taurus Molecular Cloud (as described in Sect.\,\ref{sect2}) with physical projected separations
    assuming a distance of 140\,pc. }
    \label{fig:Taurus_map}
\end{figure}

\begin{figure}
    \resizebox{\hsize}{!}{\includegraphics{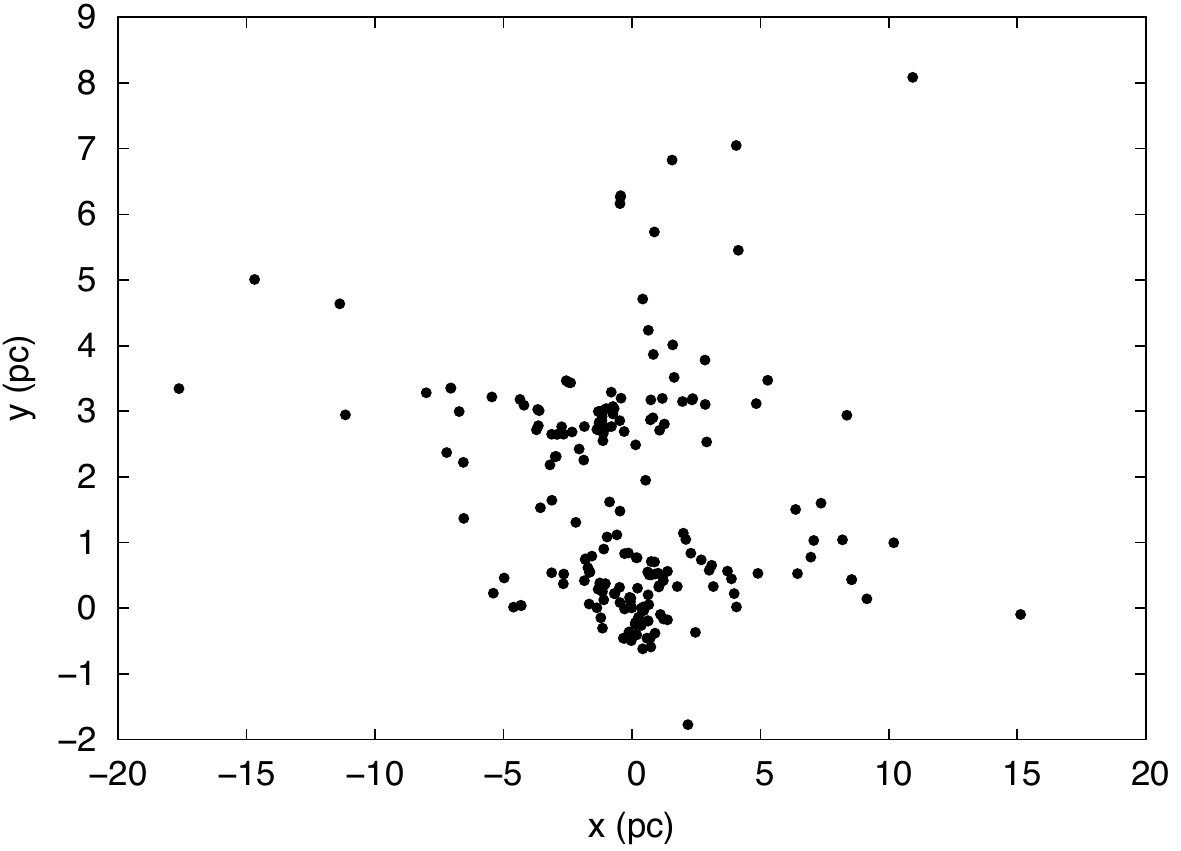}}
    \caption{The locations of each of the members of the Cham~I cluster with physical projected 
    separations assuming a distance of 160\,pc. }
    \label{fig:Cha_map}
\end{figure}

\begin{figure}
    \resizebox{\hsize}{!}{\includegraphics{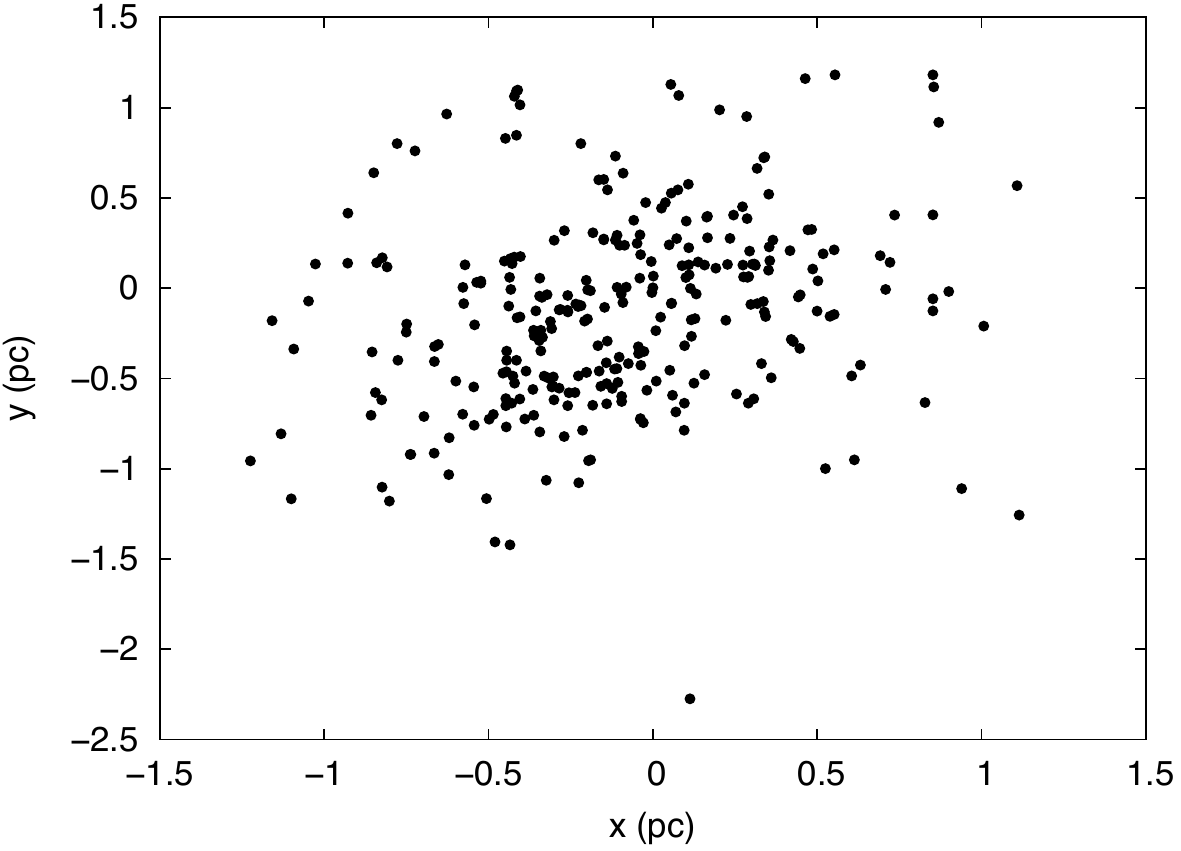}}
    \caption{The locations of each of the members of the L1688 core in Ophiuchus with physical
    projected separations assuming a distance of 130\,pc.}
    \label{fig:Oph_map}
\end{figure}

\begin{figure}
    \resizebox{\hsize}{!}{\includegraphics{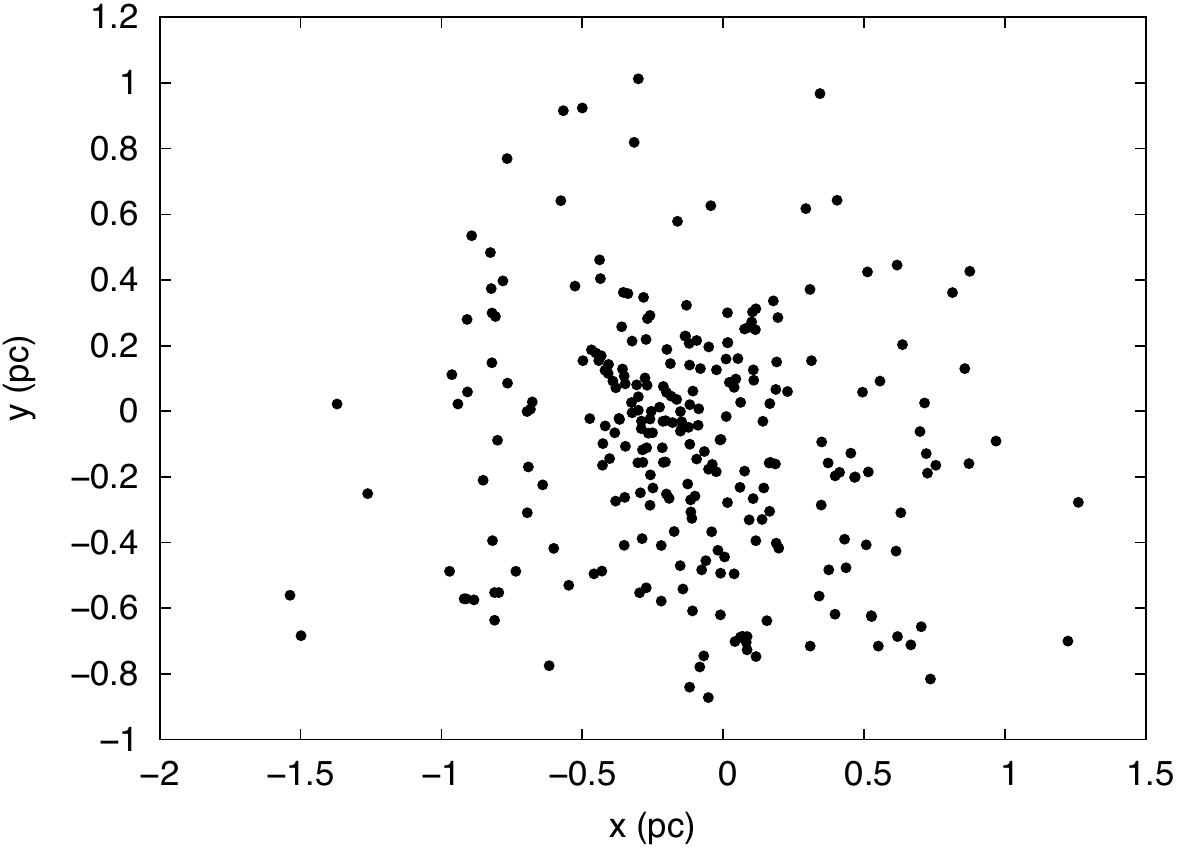}}
    \caption{The locations of each of the members of IC348 with physical projected separations
    assuming a distance of 316\,pc.}
    \label{fig:IC348_map}
\end{figure}

\begin{figure}
    \resizebox{\hsize}{!}{\includegraphics{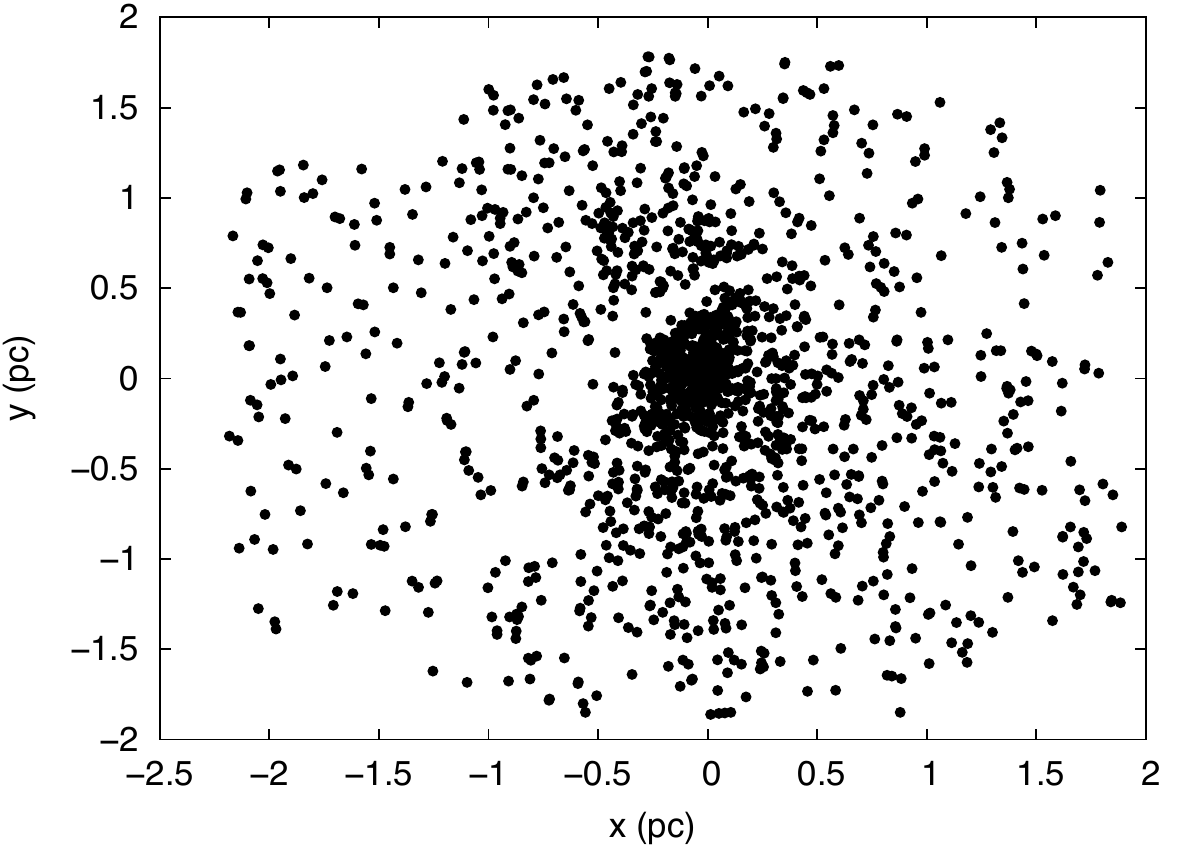}}
    \caption{The locations of each of the members of the Orion Nebula Cluster from
    \citet{Hillenbrand:1997} with physical projected separations assuming a distance of 414\,pc.
    These observations clearly miss some of the cluster members farthest from the centre and do not
    extend down to the stellar/substellar limit (see Sect.\,\ref{sect:onc}). The absorption feature 
    associated with the `lip' of the emission nebula is seen as an under-density of sources in a 
    strip across the map \citep[see][]{Hillenbrand:1997}.}
    \label{fig:ONC_map}
\end{figure}

In Figs.\,\ref{fig:Taurus_map} to \ref{fig:ONC_map} we show the (two-dimensional) stellar
distributions in each of our regions. Ophiuchus, IC348 and the ONC (Figs.~\ref{fig:Oph_map},
\ref{fig:IC348_map}, and~\ref{fig:ONC_map}) are fair examples of classic `clusters':
centrally concentrated with density declining with radius.  However, Taurus and Cham~I
(Figs.~\ref{fig:Taurus_map} and~\ref{fig:Cha_map}) are clearly very sub-structured and far
from smooth.  Despite our earlier attempt to define a `stellar density', it is very unclear
from these figures if a global `density' has any meaning whatsoever.  We address this
question in the next section and in the discussion.

\section{Simulations of binary destruction}

In this section we attempt to use the observations of the cluster morphologies and multiplicity
fractions to construct $N$-body models of star clusters to `reverse engineer'
the current state to determine the primordial binary fractions \citep{Kroupa:1995a}.

It is well-known that many binaries are destroyed in dense environments
\citep{Heggie:1975,Hills:1975}, and much theoretical work has gone into modelling the
evolution of stellar binary properties in different clustered environments. The first
comprehensive simulations were performed by \citet{Kroupa:1995b,Kroupa:1995a,Kroupa:1995c},
who showed that a primordial binary separation distribution similar to that observed in
Taurus--Auriga \citep{Leinert:1993} can evolve into a field-like \citep{Duquennoy:1991}
separation distribution if the cluster is dense enough.

\citet{Kroupa:1995b,Kroupa:1995a} and \citet{Kroupa:2011} derive a {\em universal} pre-main
sequence separation distribution based on these simulations, which is characterised by an
initial binary fraction and an excess of binaries with separations $a > 10^3$\,au, compared
to the Galactic field. Recently, \citet*{Marks:2011} have developed an analytical operator,
which, depending on the cluster's initial density, can be used to predict the effects of
dynamical evolution on the binary separation distribution in any star forming environment,
if the primordial binary population is described by the \citep{Kroupa:1995b} distribution.
This operator assumes that the cluster is roughly spherical and relaxed at birth.

However, the assumption that clusters are roughly spherical and relaxed at birth is almost
certainly not a reasonable assumption (see the distributions of Taurus and Cham~I in
Figs.~\ref{fig:Taurus_map} and~\ref{fig:Cha_map} above).  Also, although L1688 (Oph), IC348 and
the ONC appear fairly smooth now, they may well have formed in a clumpy, complex
distribution with their current smooth appearance due to dynamical evolution
\citep[see][]{Allison:2009b,Allison:2010,Parker:2011c}.  \citet{Parker:2011c} show that there
can be significant binary processing even in low-density clusters if they are initially
sub-structured.  This is because the {\em local} density can be high enough to destroy
binaries even if the average global density is very low.  After a crossing time the initial
substructure is erased and the cluster is roughly spherical and relaxed.  Therefore two
clusters that are almost identical at 1--2\,Myr old can have a very different past
dynamical history and therefore very different processing of their initial binary
populations.

In this section we will simulate the evolution of clusters starting from clumpy and smooth
initial distributions with sizes and densities chosen to roughly match our five observed
clusters.  We model the initial binary populations in the simulations as a
\citeauthor{Duquennoy:1991}-like wide log-normal with either a 45, 73 or $100$ per cent
initial binary fraction. 

\subsection{Cluster membership}

In order to match the observed clusters as closely as possible, we use approximately the
same numbers of stars in our simulations of each cluster as are observed. Therefore, our
clusters designed to mimic Cham~I contain 200 stars, IC348-like clusters contain 260
stars, the ONC-like clusters contain 1500 stars; and the Ophiuchus- and Taurus-like clusters
contain 300 stars.

We keep the number of stars fixed; however, for different primordial binary fractions this
results in different numbers of systems. For example, a Taurus-like cluster with 300 stars
and a primordial binary fraction of 100 per cent contains 150 stellar systems, all of which
are binaries; a similar cluster with a field-like binary fraction ($\sim$ 45\,per cent)
contains 200 systems, 100 of which are binaries. We note that this may underestimate the
number of stars in each cluster, as binary systems outside of the observable separation
ranges would be either unresolved or seen as two independent stars.

\subsection{Stellar systems}

To create a stellar system, the mass of the primary star is chosen randomly from a
\citet{Kroupa:2002} IMF of the form
\begin{equation}
 N(M)   \propto  \left\{ \begin{array}{ll} M^{-1.3} \hspace{0.4cm} m_0
  < M/{\rm M_\odot} \leq m_1   \,, \\ M^{-2.3} \hspace{0.4cm} m_1 <
  M/{\rm M_\odot} \leq m_2   \,,
\end{array} \right.
\end{equation}
where $m_0$ = 0.1\,M$_\odot$, $m_1$ = 0.5\,M$_\odot$, and  $m_2$ = 50\,M$_\odot$.  We do not
include brown dwarfs in the simulations as these have been removed from the observational
samples.

We then assign a secondary component to the system depending on the binary fraction
associated with the primary mass: field-like, 73 per cent, and 100 per cent.

For a field-like binary fraction we divide primaries into
four groups. Primary masses in the range 0.1~$\leq m_p/{\rm M}_\odot~<$~0.47 are M-dwarfs,
with a binary fraction of 0.42 \citep{Fischer:1992}. K-dwarfs have masses in  the range
0.47~$\leq~m_p/{\rm M}_\odot$~$<$~0.84 with a binary fraction of 0.45 \citep{Mayor:1992},
and G-dwarfs have masses from 0.84~$\leq~m_p/{\rm M}_\odot~<$~1.2  with a binary fraction of
0.57 \citep{Duquennoy:1991,Raghavan:2010}. All stars more massive than  1.2\,M$_\odot$ are
grouped together and assigned a binary fraction of unity, as massive stars have a much
larger binary fraction than low-mass stars \citep[e.g.][and references
therein]{Abt:1990,Mason:1998,Kouwenhoven:2005,Kouwenhoven:2007,Pfalzner:2007,Mason:2009}.

For the 100 per cent binary fractions {\em all} stars are in binaries.  For the 73 per cent
binary fractions, 73 per cent of all stars (regardless of mass) are in binary systems.  This
sounds like a rather arbitrary number, but as we will describe below it is this
binary fraction that provides the best fit to all of the clusters.

Secondary masses are drawn from a flat mass ratio distribution; recent work by
\citet{Reggiani:2011} has shown the mass ratio of field binaries to be consistent with being
drawn from a flat distribution, rather than random pairing from the IMF.

We draw the periods of the binary systems from the log$_{10}$-normal fit to the G-dwarfs in
the field by \citet[][]{Duquennoy:1991} -- see also \citet{Raghavan:2010}, which has also
been extrapolated to fit the period distributions of the K- and M-dwarfs
\citep{Mayor:1992,Fischer:1992}:
\begin{equation}
f\left({\rm log_{10}}P\right)  \propto {\rm exp}\left \{ \frac{-{({\rm log_{10}}P -
\overline{{\rm log_{10}}P})}^2}{2\sigma^2_{{\rm log_{10}}P}}\right \},
\end{equation}
where $\overline{{\rm log_{10}}P} = 4.8$, $\sigma_{{\rm log_{10}}P} = 2.3$ and $P$ is  in
days. We convert the periods to semi-major axes using the masses of the binary components.

The eccentricities of binary stars are drawn from a thermal distribution
\citep{Heggie:1975,Kroupa:2008} of the form
\begin{equation}
f_e(e) = 2e.
\end{equation}
In the sample of \citet{Duquennoy:1991}, close binaries (with periods less than 10 days) are
almost exclusively on tidally circularised orbits. We account for this by reselecting the
eccentricity of a system if it exceeds the following period-dependent
value \citep{Parker:2011b}:
\begin{equation}
e_{\rm tid} = \frac{1}{2}\left[0.95 + {\rm tanh}\left(0.6\,{\rm log_{10}}P - 1.7\right)\right].
\end{equation}

We combine the primary and secondary masses of the binaries with their semi-major axes and
eccentricities to determine the relative velocity and radial components of the stars in each
system. The binaries are then placed at the centre of mass and velocity for each system in a
fractal or Plummer sphere (see below).

\subsection{Numerical parameters}

The simulations are run for 10\,Myr using the \texttt{kira} integrator in the Starlab
package \citep[e.g.][]{Zwart:1999,Zwart:2001} and the binary fractions and densities are
determined after 1\,Myr. We do not include stellar evolution in the simulations. As no
systems of higher order than $n=2$ form, the binary fraction is equivalent to the
multiplicity fraction. Details of each simulation are presented in
Table~\ref{cluster_props}.

We determine whether a star is in a bound binary system using the nearest neighbour
algorithm outlined in \citet{Parker:2009} and \citet{Kouwenhoven:2010}. If two stars are
closer than the average local stellar separation, are also mutual nearest neighbours, and have
a negative binding energy, then they are in a bound binary system.

In principle, this differs from an observer's definition of a visual binary, which could
include chance associations along the line of sight. However, numerical experiments indicate
that the total number that could merely be chance associations is negligible.

\subsection{`Observing' simulations}

We analyse the binary fractions of clusters in a way as close as possible to the
observations.  We only `observe' binaries in the separation ranges matched by the real
observations, taking closer binaries to be single unresolved stars, and wider binaries to be
two separate stars.  We only `observe' systems with primary masses in the range
0.1\,M$_\odot \leq m_p < $ 3.0\,M$_\odot$, and with mass ratios $q = m_s/m_p \geq 0.1$.

To determine the stellar densities of the clusters we use the same method as applied to the
real clusters, determining the volume densities within 0.25~pc from a
2D centroid fit to the centre of the cluster for our models for the
\mbox{ONC-,} Ophiuchus- and IC348-like clusters. As with the observations, such a determination is problematic for
sub-structured distributions, but we use it in the absence of anything
better. 

For our Cham~I-like and Taurus-like clusters, we measure the
stellar surface density for each star, according to the prescription
of \citet{Casertano:1985}:
\begin{equation}
\Sigma = \frac{N - 1}{\pi D^2_N},
\end{equation}
where $N$ is the $N^{\rm th}$ nearest neighbour (we choose $N = 7$) and $D_N$ is the
projected distance to that nearest neighbour. We then determine the
star with the highest surface density and measure the volume density
from that star, adopting a radius of 0.25~pc for Cham~I and 1~pc
for Taurus. 

\subsection{Cluster set-up and morphologies}

We assume two different morphologies for the initial conditions of our clusters. Firstly, we
model clusters as radially smooth Plummer spheres \citep{Plummer:1911}, which are used
extensively in modelling the dynamical evolution of star clusters
\citep[e.g.][]{Kroupa:1999,Moraux:2007,Parker:2009}. Secondly, we adopt a fractal
distribution so that our clusters contain substructure
initially \citep[e.g.][]{Goodwin:2004a,Allison:2010,Parker:2011c}. In two dimensions these
model set-ups reproduce, to first order, the entire range of observed morphologies described
in Section~5, although we note that other set-ups, such as the King profile
\citep[e.g.][]{King:1966} can, and have been used to fit several of the observed clusters
\citep[e.g. the ONC][]{Hillenbrand:1998}.

Our aim is to produce clusters that have the same morphology, density,
and binary fractions as the observed clusters at the age of the
observed clusters.

\begin{figure*}
 \begin{center}
\setlength{\subfigcapskip}{10pt}
%\hspace*{-1.5cm}
\subfigure[Plummer sphere, 0\,Myr]{\label{ONC-morph-a}{\includegraphics[scale=0.31]{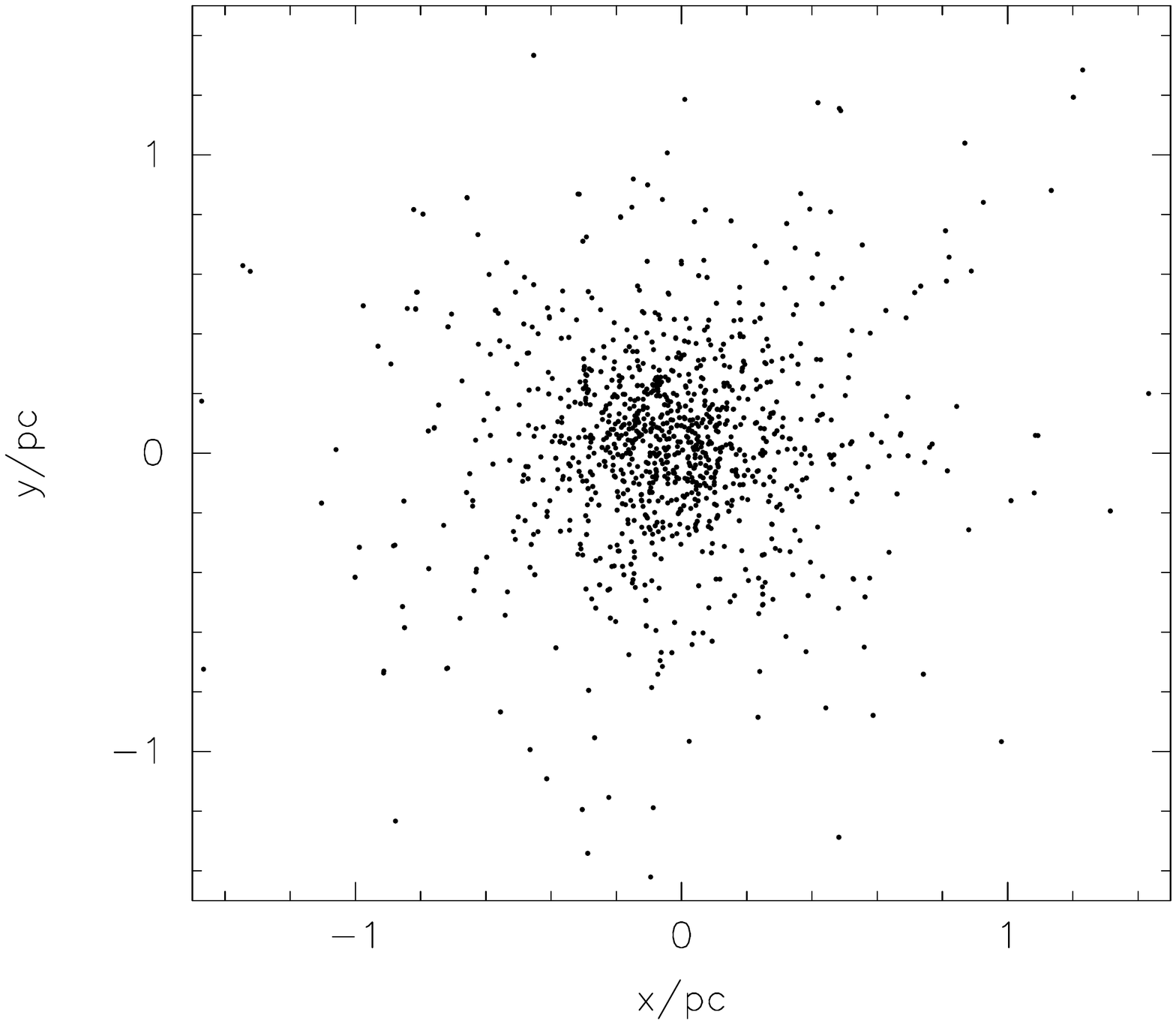}}}
\hspace*{-0.3cm} 
\subfigure[Plummer sphere, 1\,Myr]{\label{ONC-morph-b}{\includegraphics[scale=0.31]{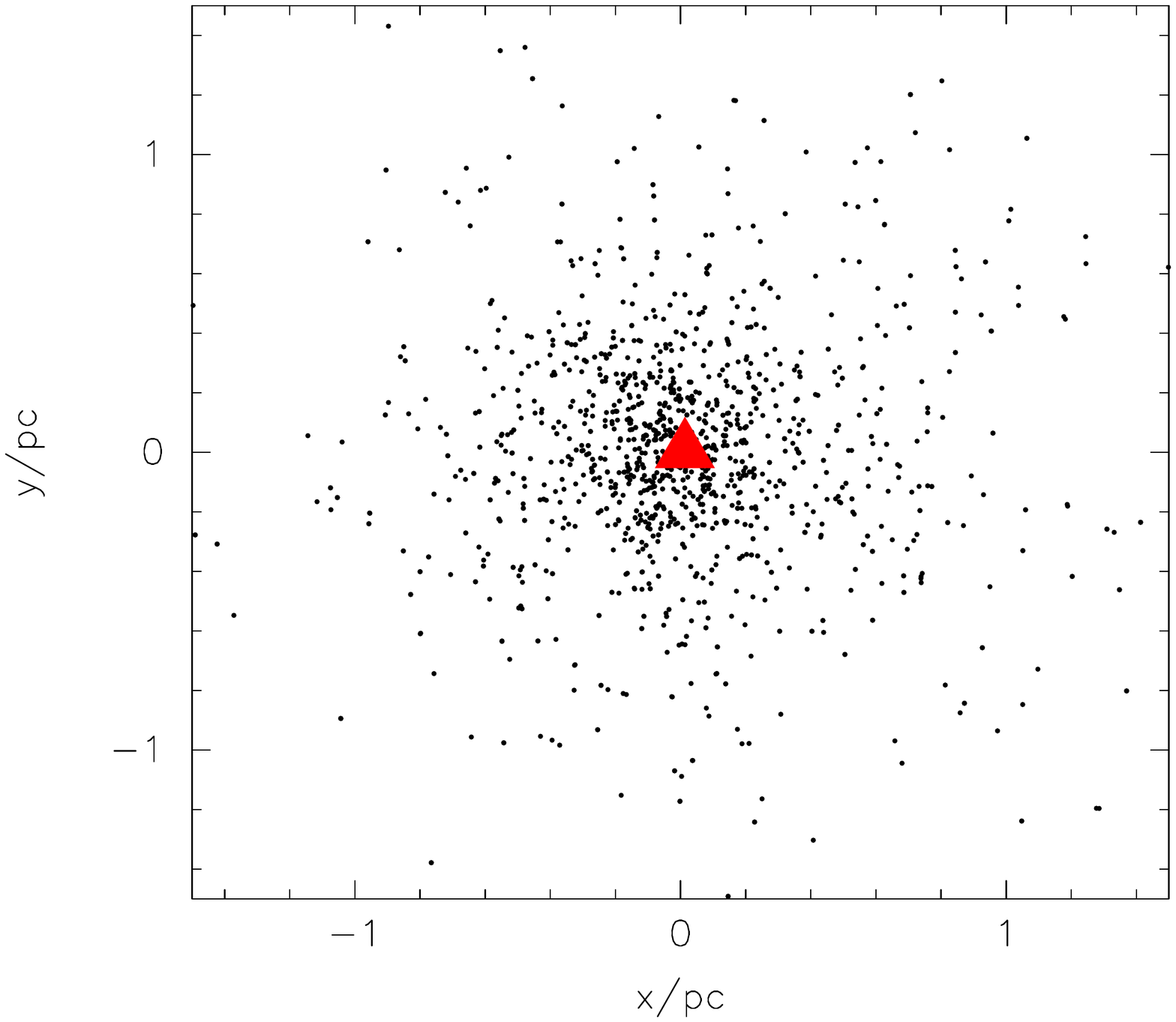}}} 
%\hspace*{-1.5cm}
\subfigure[Fractal, 0\,Myr]{\label{ONC-morph-c}{\includegraphics[scale=0.31]{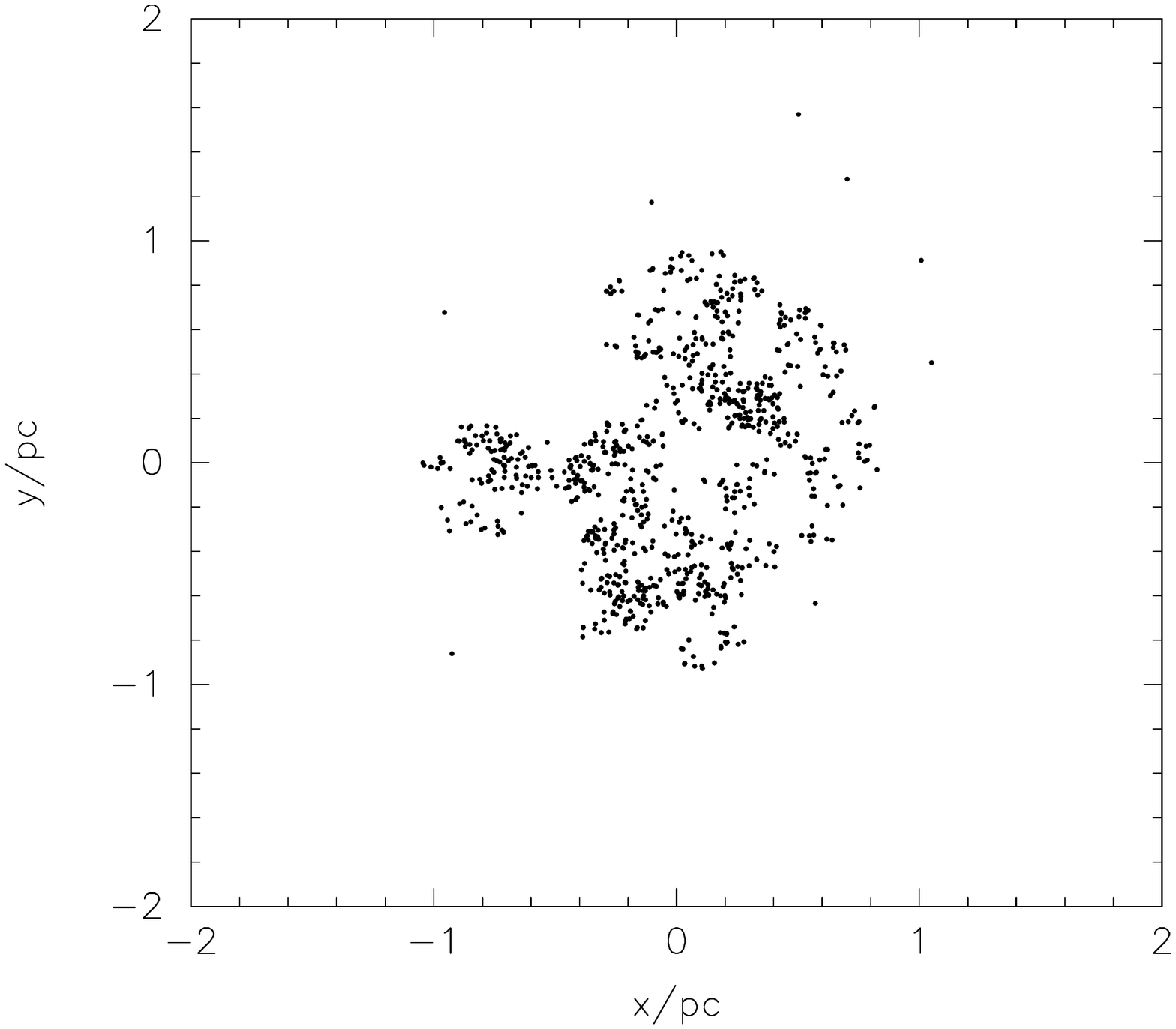}}}
\hspace*{-0.3cm} 
\subfigure[Fractal, 1\,Myr]{\label{ONC-morph-d}{\includegraphics[scale=0.31]{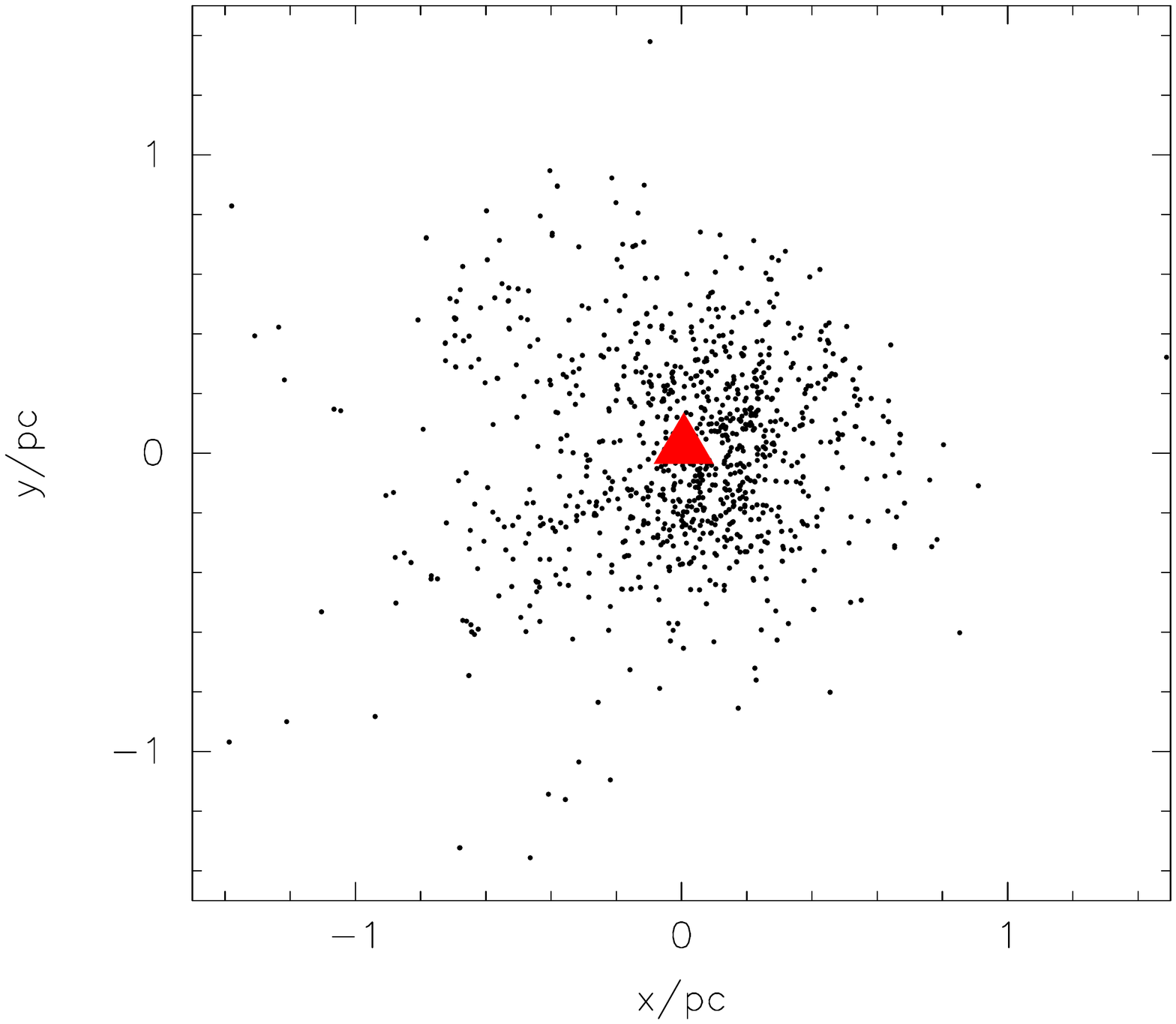}}} 
\caption[bf]{Typical morphologies for our different initial conditions for ONC-like clusters. In the top panels we show a Plummer sphere with an initial half-mass radius 
of 0.4\,pc at (a) 0\,Myr and (b) 1\,Myr. In the bottom panels we show a collapsing fractal cluster at (c) 0\,Myr and (d) 1\,Myr. 
The centroid position in the cluster at 1\,Myr is marked in panels (b) and (d) by the red triangle.}
\label{ONC_morph}
\end{center}
\end{figure*}

\begin{figure*}
 \begin{center}
\setlength{\subfigcapskip}{10pt}
%\hspace*{-1.5cm}
\subfigure[Plummer sphere, 0\,Myr]{\label{IC348-morph-a}{\includegraphics[scale=0.31]{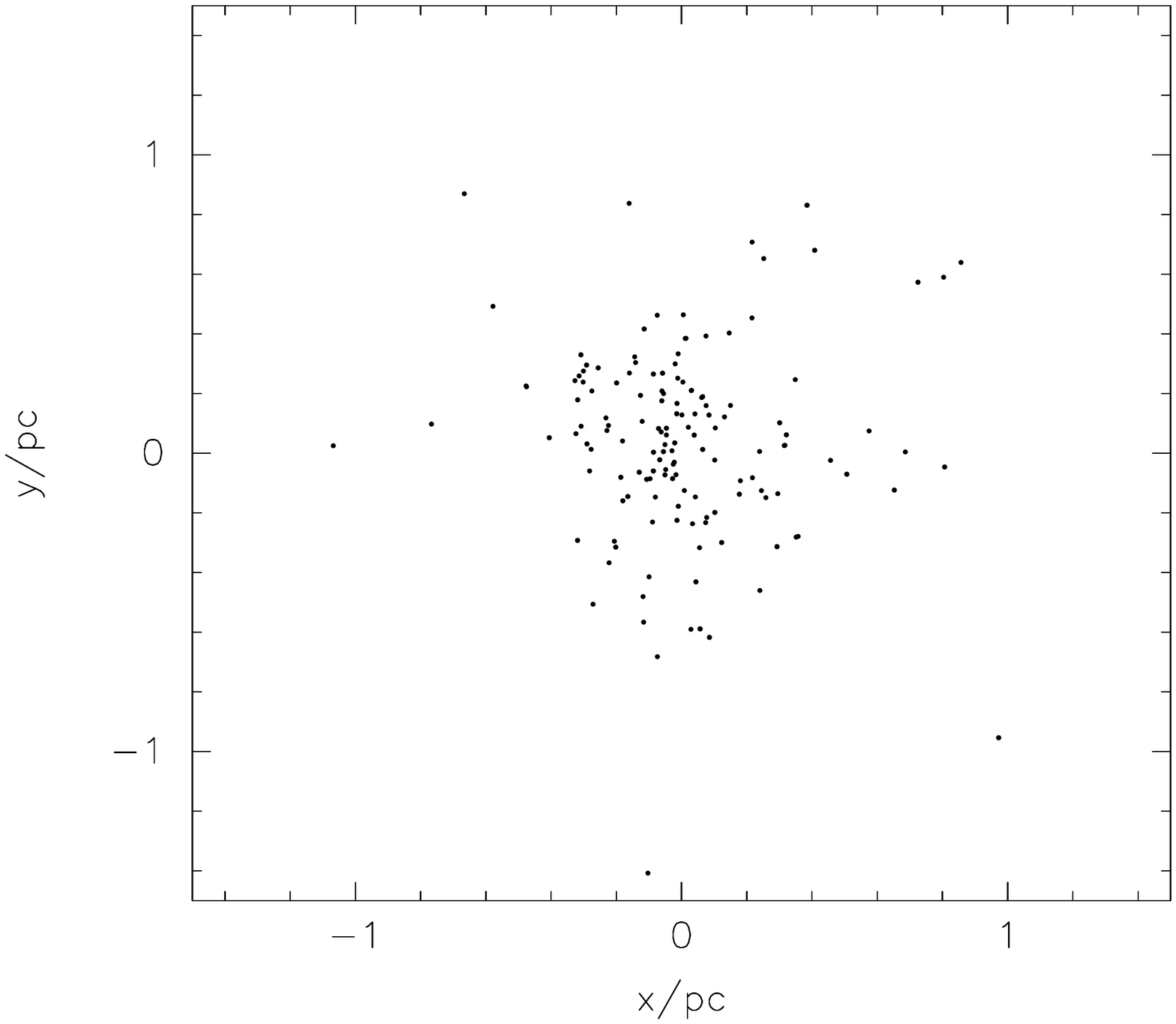}}}
\hspace*{-0.3cm} 
\subfigure[Plummer sphere, 1\,Myr]{\label{IC348-morph-b}{\includegraphics[scale=0.31]{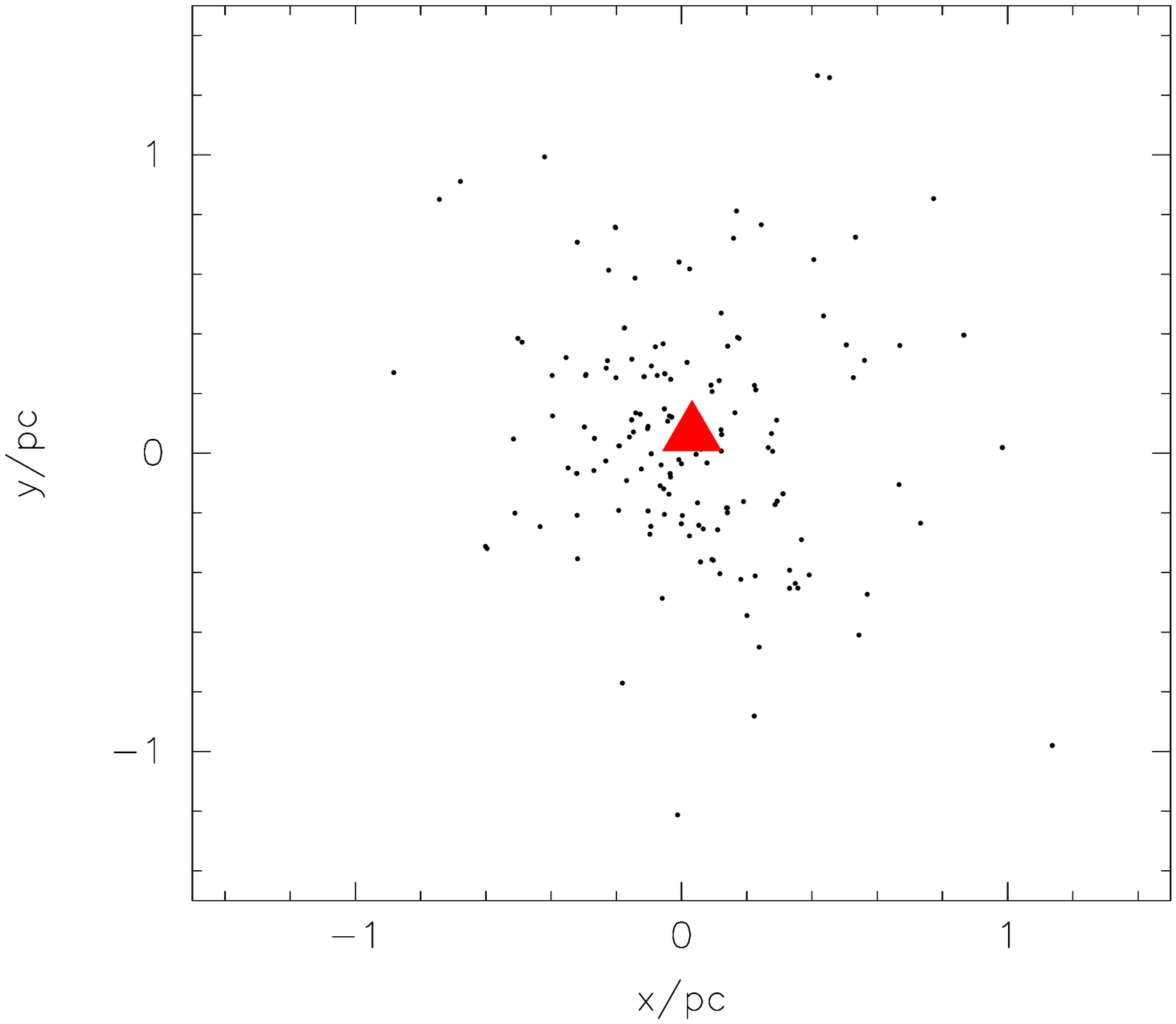}}} 
%\hspace*{-1.5cm}
\subfigure[Fractal, 0\,Myr]{\label{IC348-morph-c}{\includegraphics[scale=0.31]{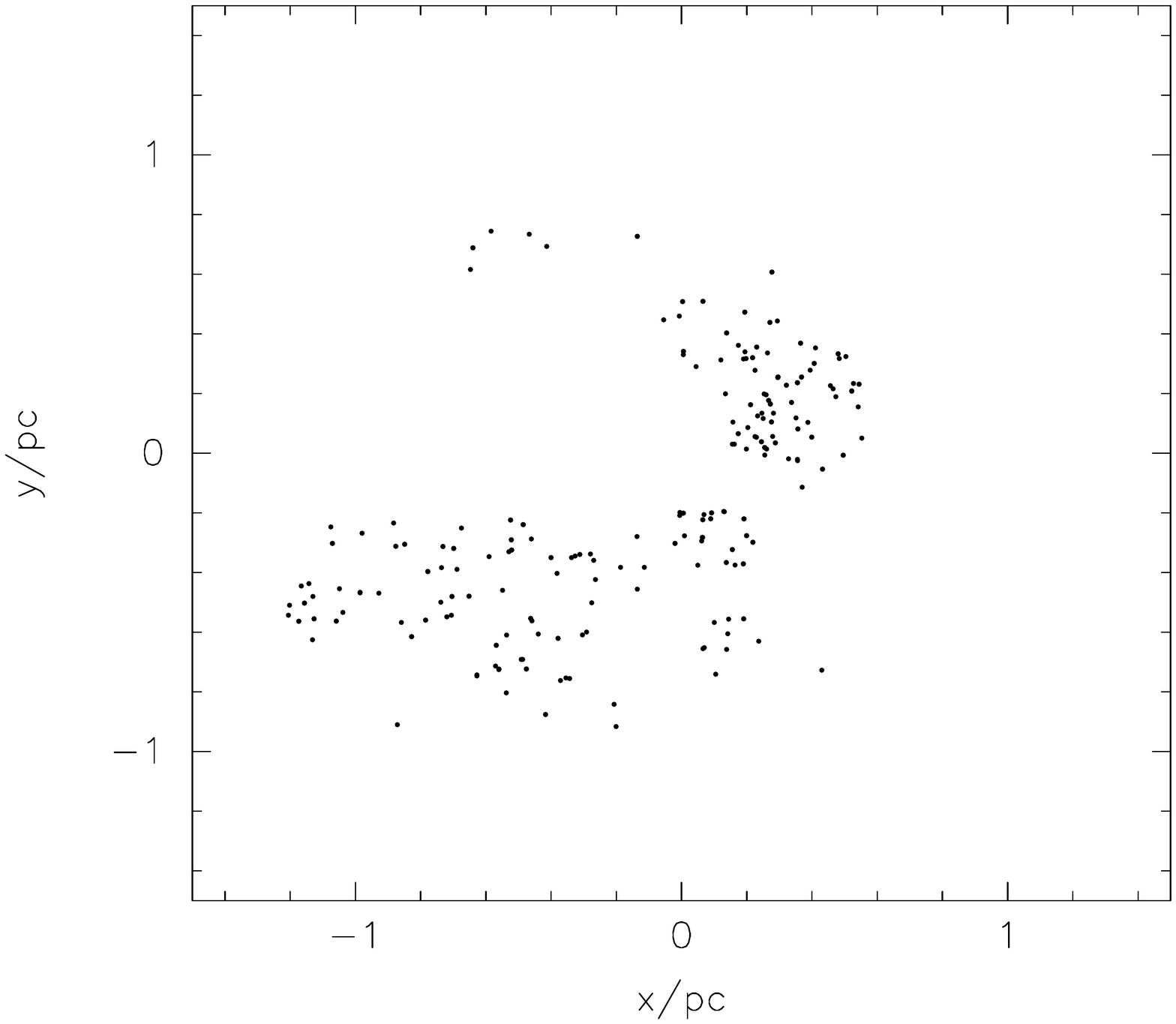}}}
\hspace*{-0.3cm} 
\subfigure[Fractal, 1\,Myr]{\label{IC348-morph-d}{\includegraphics[scale=0.31]{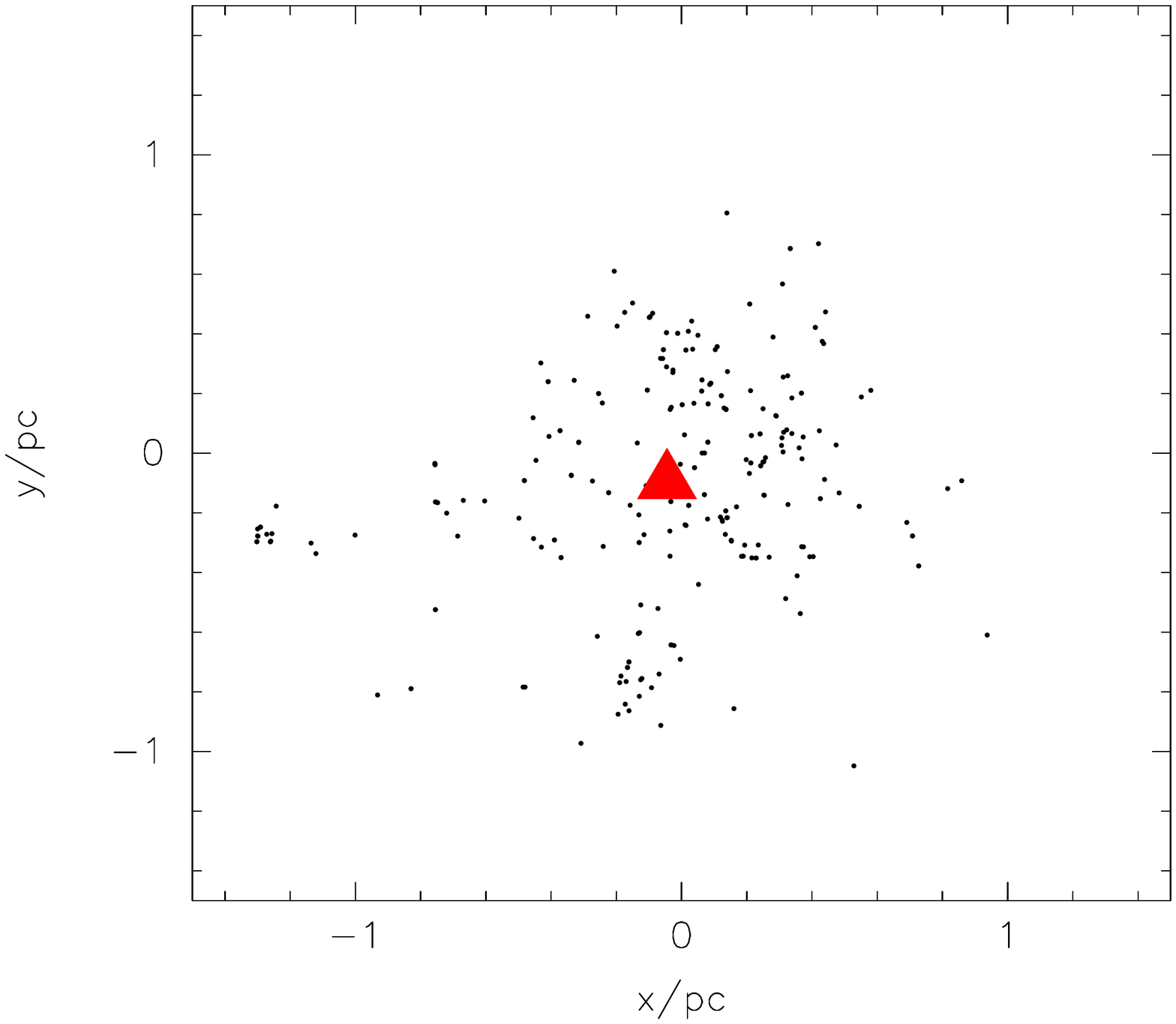}}} 
\caption[bf]{Typical morphologies for our different initial conditions for IC348-like clusters. In the top panels we show a Plummer sphere with an initial half-mass radius 
of 0.4\,pc at (a) 0\,Myr and (b) 1\,Myr. In the bottom panels we show a collapsing fractal cluster at (c) 0\,Myr and (d) 1\,Myr. 
The centroid position in the cluster at 1\,Myr is marked in panels (b) and (d) by the red triangle.}
\label{IC348_morph}
\end{center}
\end{figure*}

\begin{figure*}
 \begin{center}
\setlength{\subfigcapskip}{10pt}
%\hspace*{-1.5cm}
\subfigure[Plummer sphere, 0\,Myr]{\label{Oph-morph-a}{\includegraphics[scale=0.31]{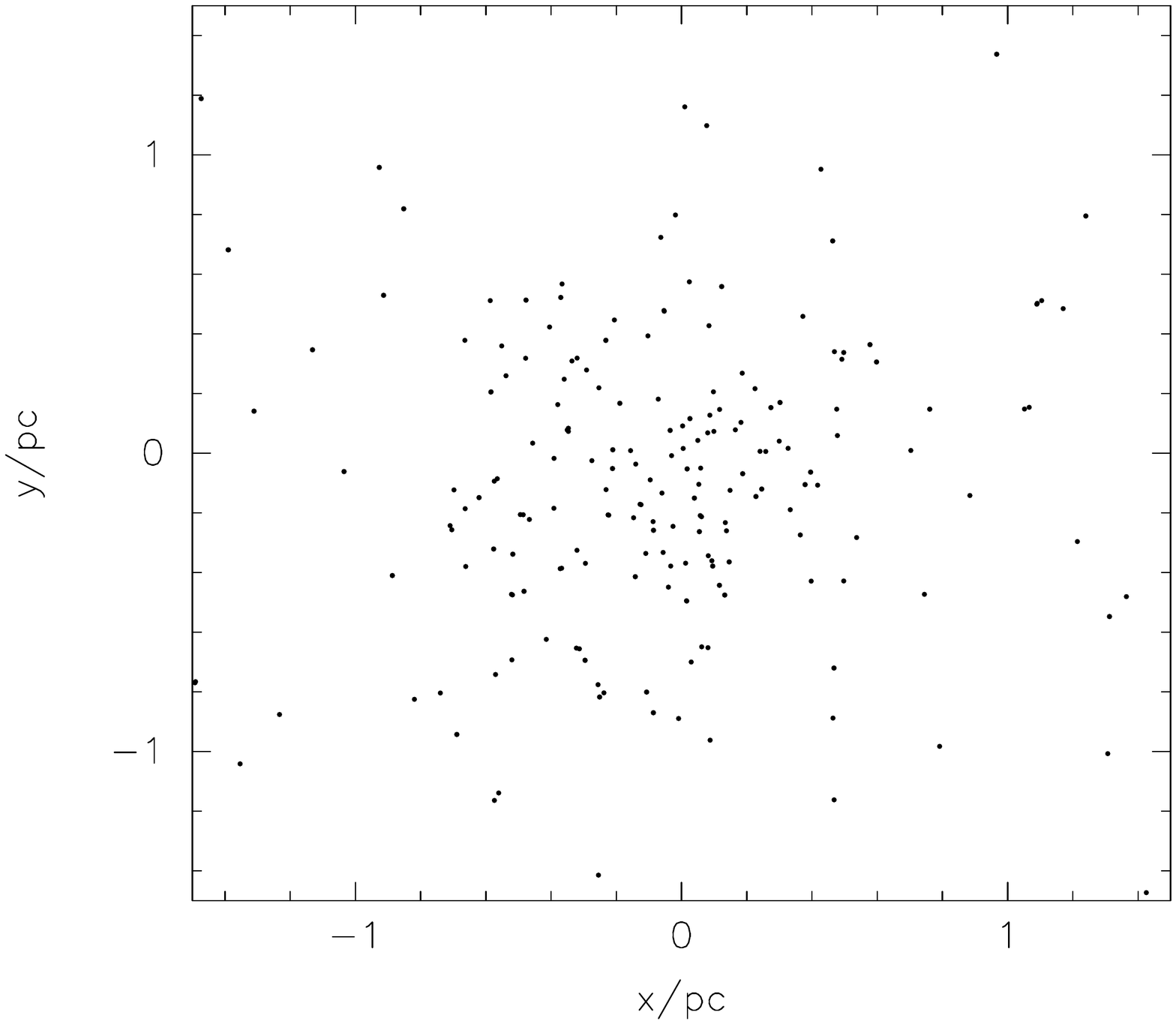}}}
\hspace*{-0.3cm} 
\subfigure[Plummer sphere, 1\,Myr]{\label{Oph-morph-b}{\includegraphics[scale=0.31]{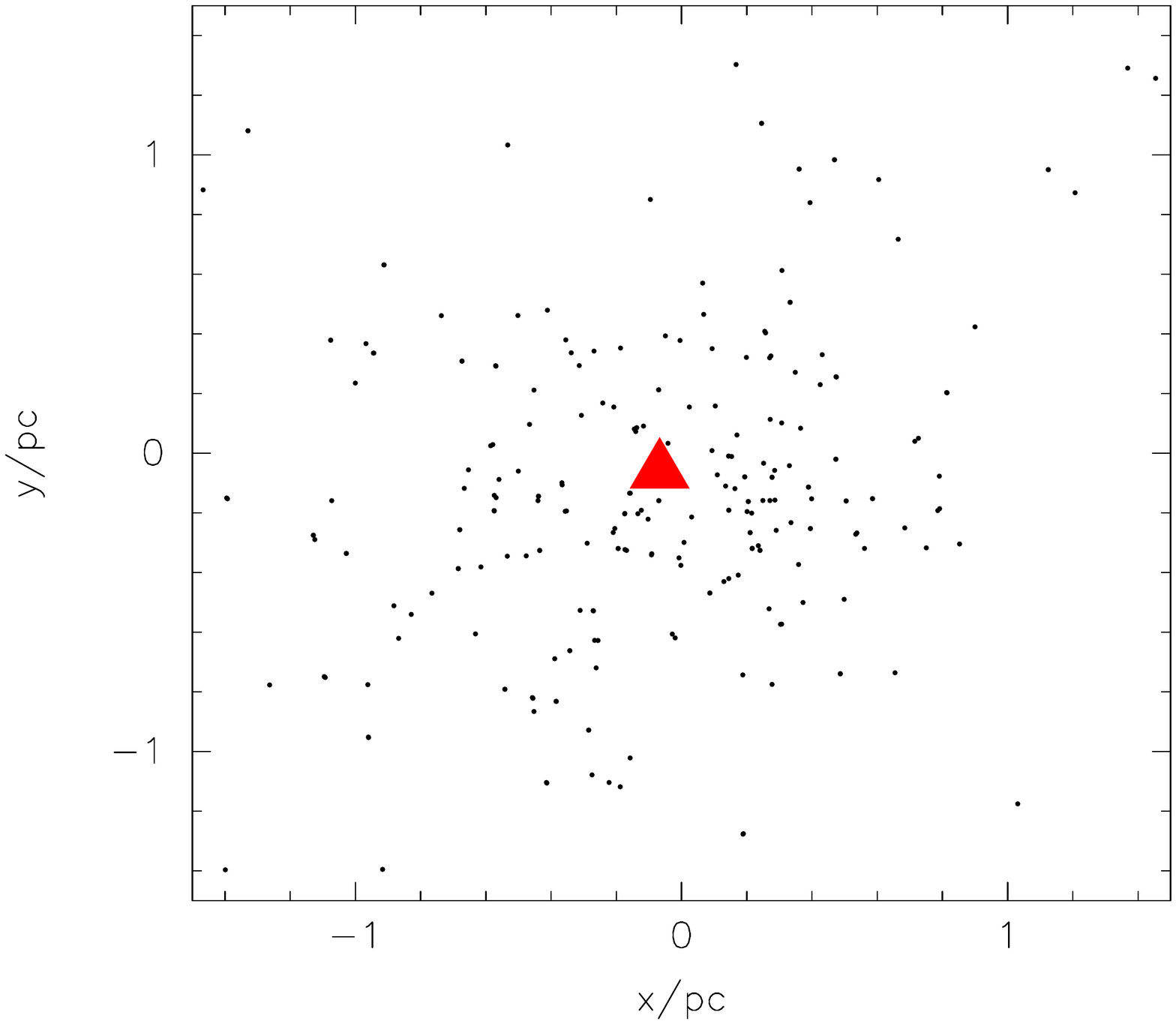}}} 
%\hspace*{-1.5cm}
\subfigure[Fractal, 0\,Myr]{\label{Oph-morph-c}{\includegraphics[scale=0.31]{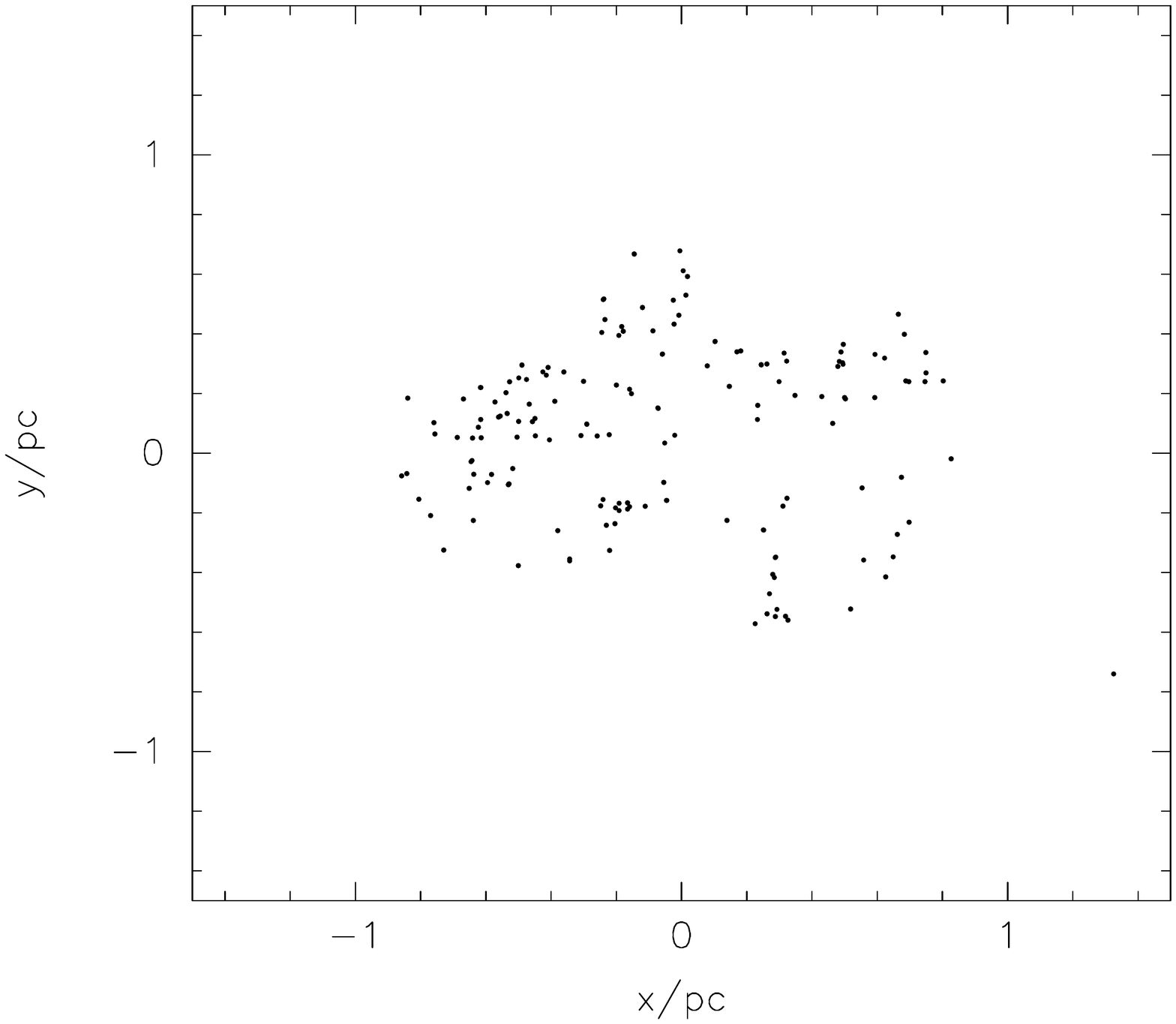}}}
\hspace*{-0.3cm} 
\subfigure[Fractal, 1\,Myr]{\label{Oph-morph-d}{\includegraphics[scale=0.31]{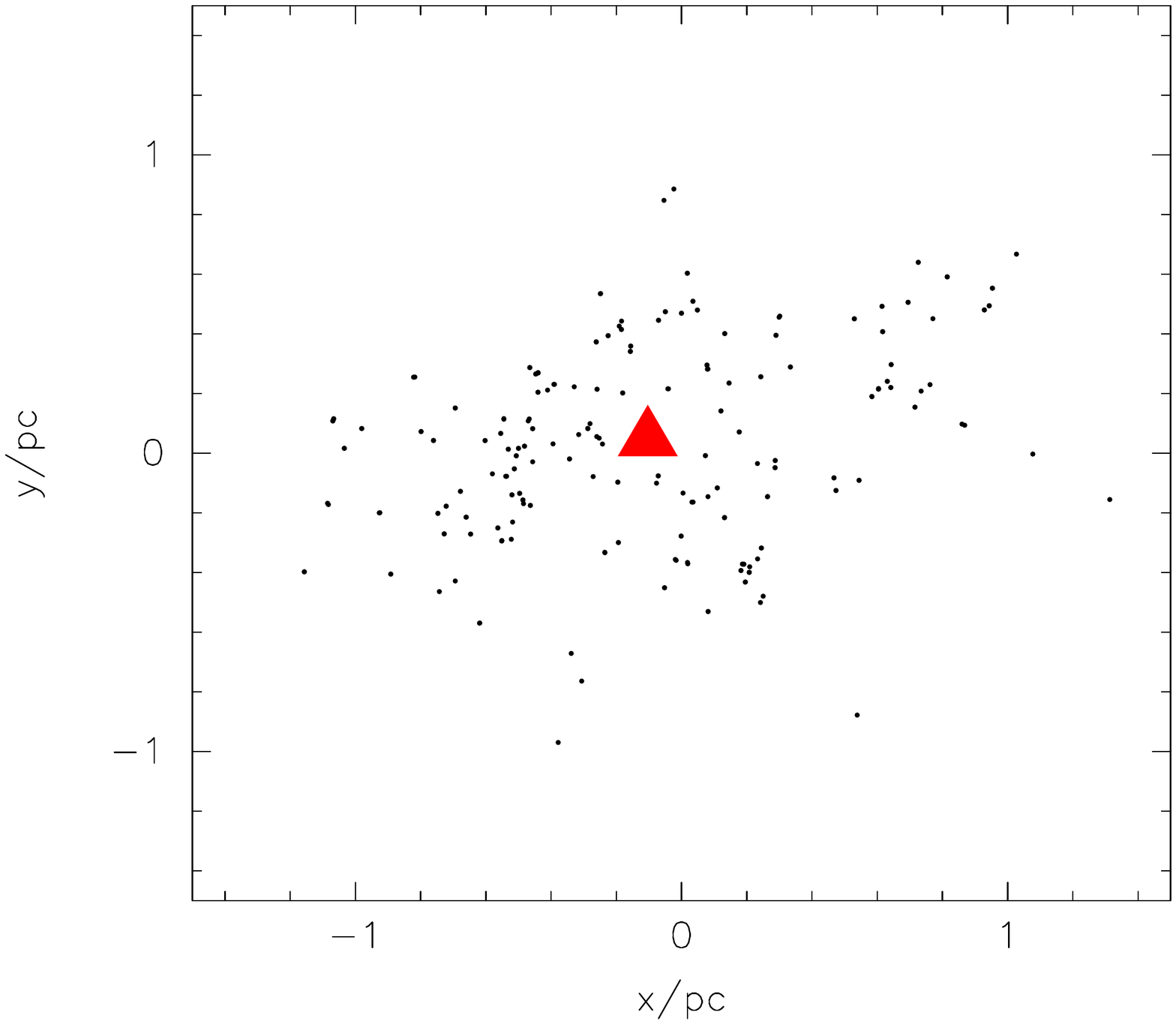}}} 
\caption[bf]{Typical morphologies for our different initial conditions for Ophiuchus-like clusters. In the top panels we show a Plummer sphere with an initial half-mass radius 
of 0.8\,pc at (a) 0\,Myr and (b) 1\,Myr. In the bottom panels we show a fractal cluster in virial equilibrium at (c) 0\,Myr and (d) 1\,Myr. 
The centroid position in the cluster at 1\,Myr is marked in panels (b) and (d) by the red triangle.}
\label{Oph_morph}
\end{center}
\end{figure*}

\begin{figure*}
 \begin{center}
\setlength{\subfigcapskip}{10pt}
\subfigure[Fractal, 0\,Myr]{\label{Cha-morph-a}{\includegraphics[scale=0.31]{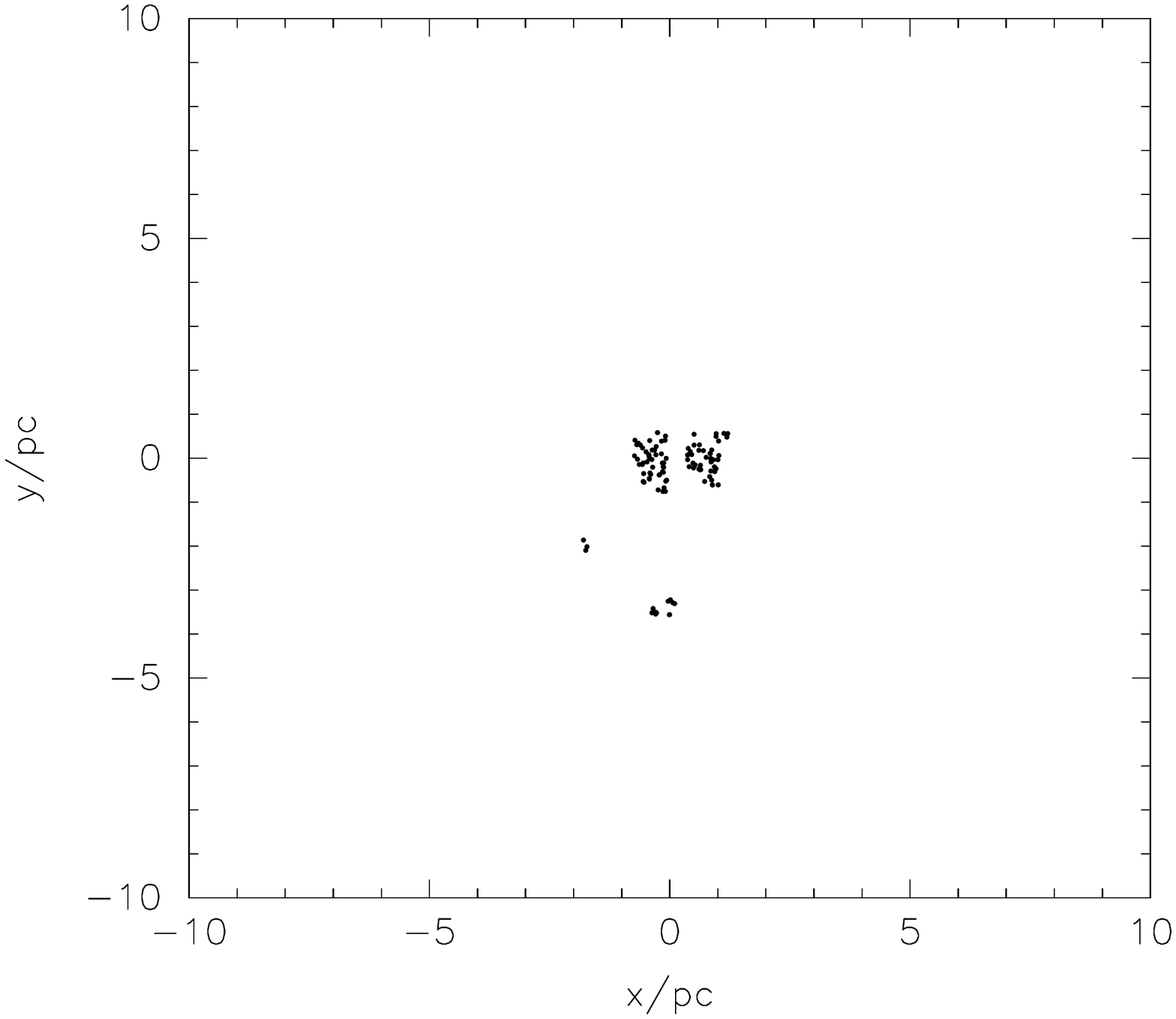}}}
\hspace*{-0.3cm} 
\subfigure[Fractal, 1\,Myr]{\label{Cha-morph-b}{\includegraphics[scale=0.31]{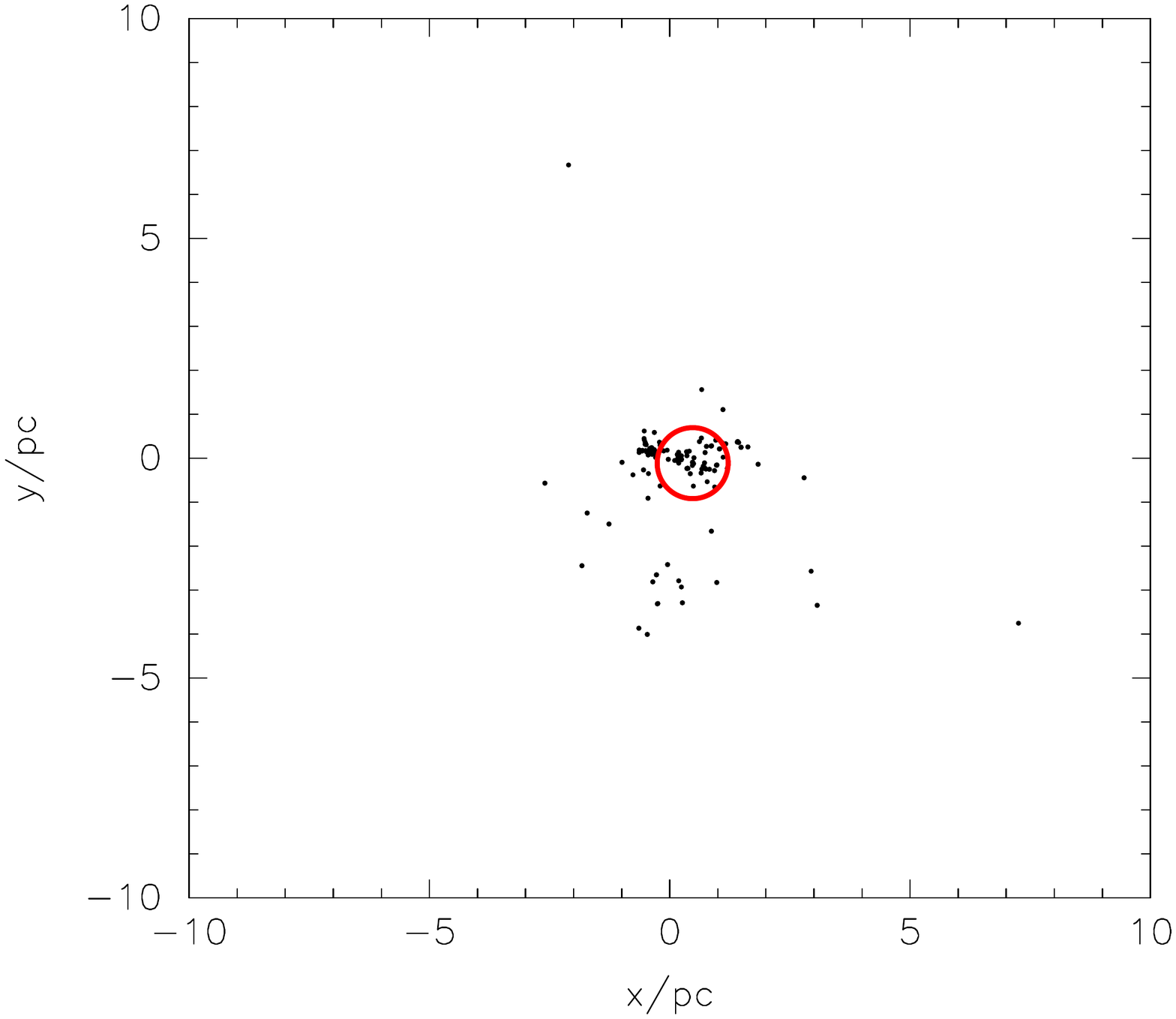}}} 
\caption[bf]{Typical morphologies for Chamaeleon-like clusters. We show a fractal in virial equilibrium at (a) 0\,Myr and (b) 1\,Myr. We show the star with the maximum local surface density 
(the point from which we measure the stellar density) at 1\,Myr by a red circle.}
\label{Cha_morph}
\end{center}
\end{figure*}

\begin{figure*}
 \begin{center}
\setlength{\subfigcapskip}{10pt}
\subfigure[Fractal, 0\,Myr]{\label{Tau-morph-a}{\includegraphics[scale=0.31]{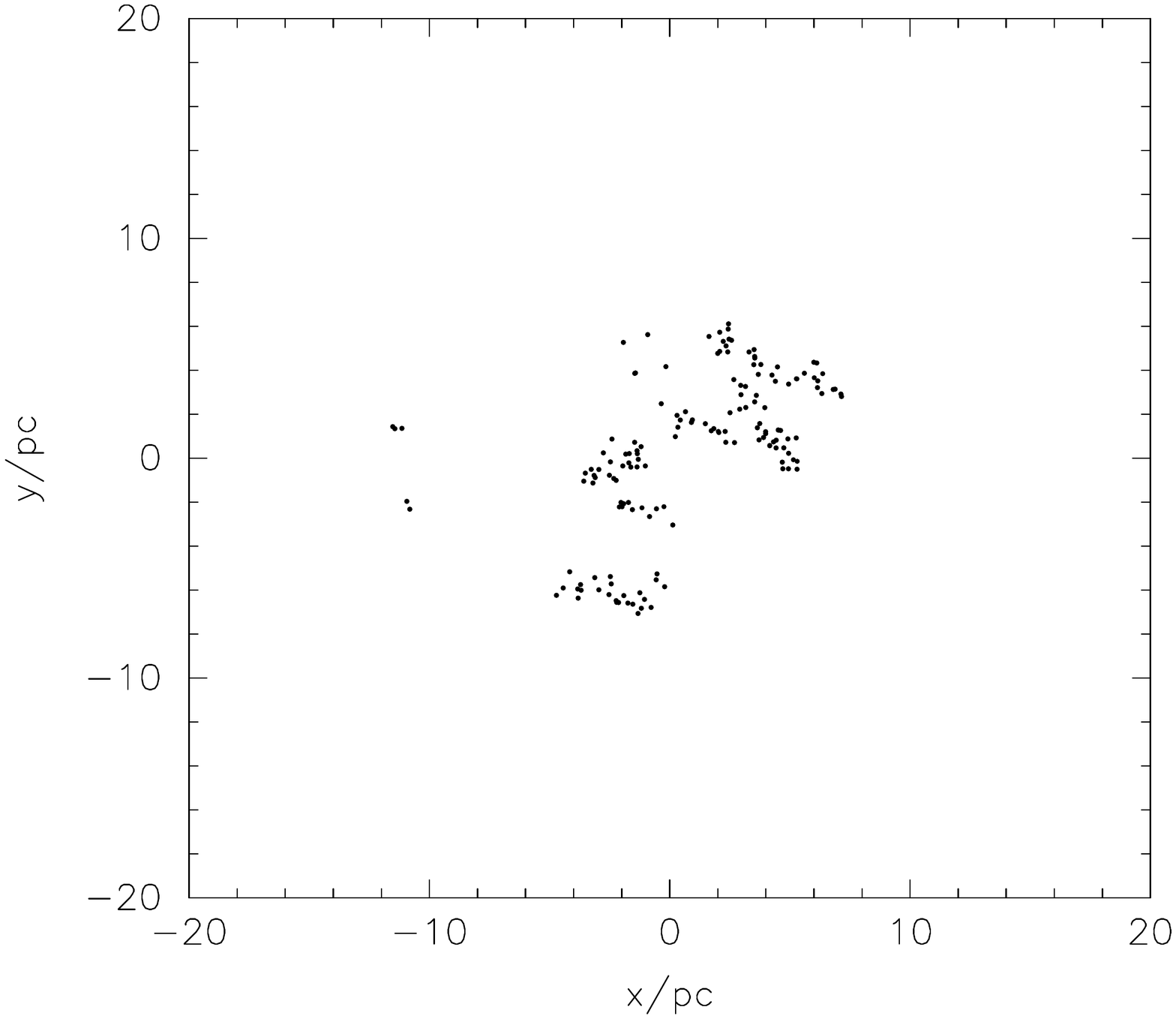}}}
\hspace*{-0.3cm} 
\subfigure[Fractal, 1\,Myr]{\label{Tau-morph-b}{\includegraphics[scale=0.31]{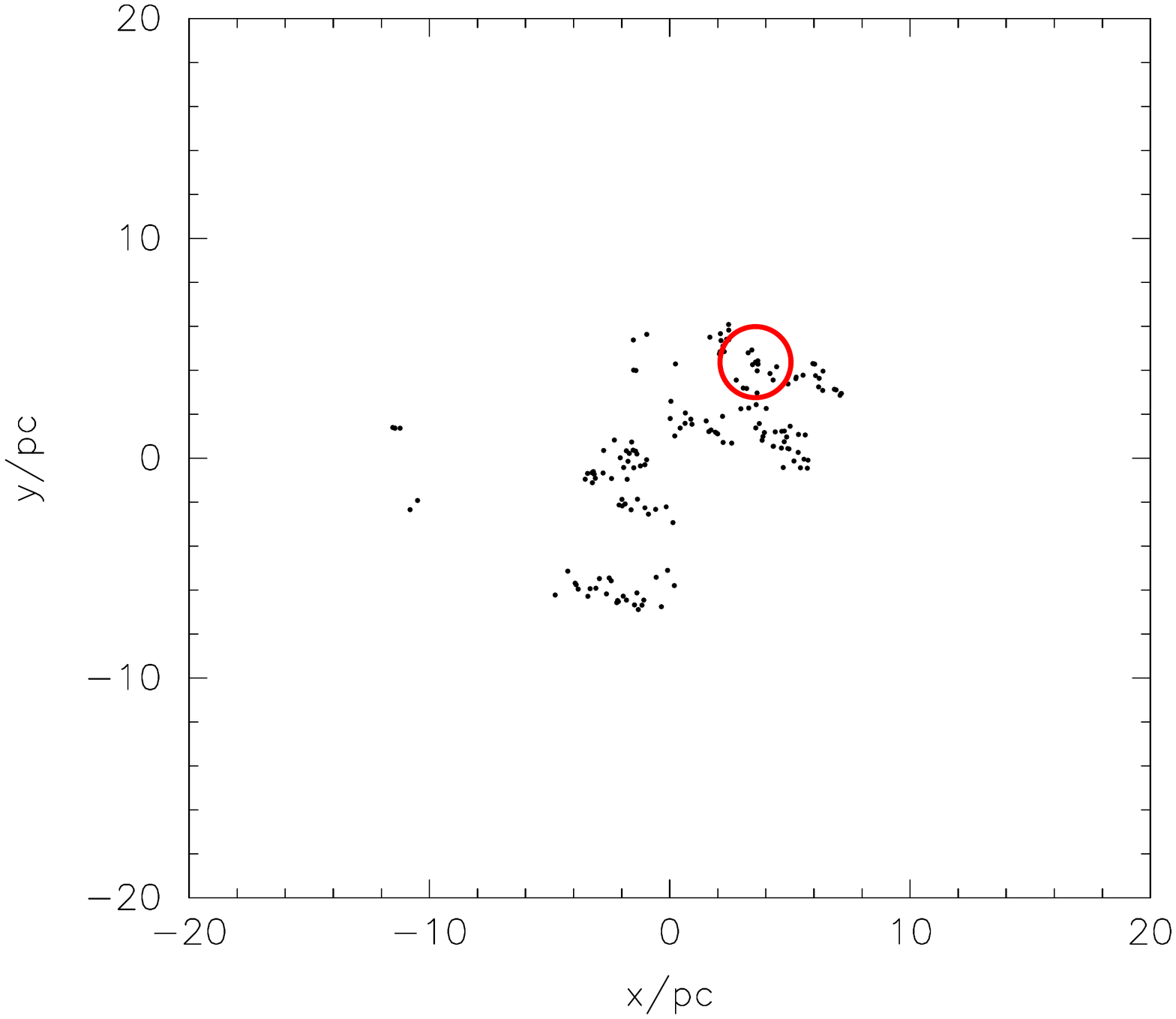}}} 
\caption[bf]{Typical morphologies for Taurus-like clusters. We show a fractal in virial equilibrium at (a) 0\,Myr and (b) 1\,Myr. We show the star with the maximum local surface density 
(the point from which we measure the stellar density) at 1\,Myr by a red circle.}
\label{Tau_morph}
\end{center}
\end{figure*}

\subsubsection{Plummer spheres}

Plummer spheres (and fractals, see below) are used as initial conditions for the clusters
that are observed to be roughly spherical, namely the ONC, Ophiuchus and IC348.  Due to
their obviously substructured nature we do not attempt to model Cham~I  or Taurus with
Plummer spheres as initial conditions. We set up Plummer spheres according to the
prescription in \citet*{Aarseth:1974}, and we assume that they are in virial equilibrium at
the start of the simulations.

We assume three different half-mass radii  for the simulated clusters; 0.1, 0.4 and 0.8\,pc.
For an ONC-like cluster, a half-mass radius of 0.1\,pc corresponds to an initial density of
$\sim 10^4$\,M$_\odot$pc$^{-3}$, which is significantly higher than the observed value.
However, \citet{Parker:2009} show that such a cluster expands during the first 1\,Myr of
evolution and may have a similar density to the ONC after this time. A half-mass radius of
0.8\,pc corresponds to the half-mass radius of the ONC today \citep{Hillenbrand:1998}.

We show examples of Plummer sphere morphologies for simulations of three of the observed
clusters (details of the cluster set-up are given in Table~\ref{cluster_props}). In
Fig.~\ref{ONC_morph} we show a Plummer sphere with 1500 stars, and a half-mass
radius\footnote{We assume that there is no primordial mass segregation, so this also
corresponds to the half-number radius.} of 0.4\,pc (simulation ID = 2 in
Table~\ref{cluster_props}, before dynamical evolution (Fig.~\ref{ONC-morph-a}) and at 1\,Myr
(Fig.~\ref{ONC-morph-b}). This corresponds to an ONC-like cluster, if the cluster formed
without substructure.

We also show Plummer sphere morphologies for our IC348- and Ophiuchus-like clusters. In
Fig.~\ref{IC348_morph} we show a Plummer sphere with 260 stars, and a half-mass radius of
0.4\,pc (IC348, simulation ID = 7 in Table~\ref{cluster_props}), and in Fig.~\ref{Oph_morph}
we show a Plummer sphere with 300 stars, and a half-mass radius of 0.8\,pc (Ophiuchus,
simulation ID = 13 in Table~\ref{cluster_props}). In both figures, panel (a) shows the
cluster before dynamical evolution, and panel (b) shows the cluster at 1\,Myr.

All the Plummer sphere clusters are in virial equilibrium ($Q = 0.5$,
where $Q = T/|\Omega|$, and $T$ and $\Omega$ are the total kinetic energy
and total potential energy of the stars) at the start of the simulations.

\subsubsection{Fractals}

Observations of young star forming regions show that a large amount of substructure is
present in young star-forming regions \citep[e.g.][]{Sanchez:2009,Schmeja:2011}. We employ a
relatively straightforward way of modelling substructure using a fractal \citep[see
e.g.,][]{Goodwin:2004a}. In a fractal, the level of substructure is set by
just one number, the fractal dimension $D$. A highly sub-structured cluster has $D = 1.6$,
whereas a cluster with no substructure has $D = 3.0$. For the ONC-like clusters we adopt $D
= 2.0$ (a moderate amount of substructure), whereas the Cham~I- and Taurus-like
clusters have $D = 1.6$.

We refer the interested reader to \citet{Goodwin:2004a,Allison:2010,Parker:2011c} for a full
description of the fractal set-up. Here, we briefly summarise the main features of the
model. A cube is constructed, at the centre of which a `parent' is placed. This then spawns
subcubes, each of which contains a child at its centre. The fractal is built by determining
how many of the children become parents, which is governed by the fractal dimension, $D$.
For a lower fractal dimension, fewer children survive, and the cluster contains more
substructure. The cube is pruned to make a sphere, and then children
are randomly removed until the required number of `stars' remain.

The velocities of stars in the fractal are determined thus; first generation children are
assigned velocities from a Gaussian of mean zero, and children inherit their parent's
velocity plus a random component that decreases with each generation in the fractal. This
results in a velocity structure where nearby stars have very similar velocities, but distant
stars can have very different velocities. The velocity of every star is then scaled to
obtain the desired virial ratio of the cluster.

We set up some fractal clusters in virial equilibrium ($Q=0.5$), and
others are sub-virial (cool, $Q = 0.3$). \citet{Allison:2010} have
shown that a $N=1000$ fractal with an initial size of 1\,pc, and
sub-virial velocities, will collapse and reach a very dense phase after $\sim$1\,Myr. This
model has been successful in matching the observed levels of mass segregation in the ONC
through dynamics \citep{Allison:2009b}, forming Trapezium-like systems
\citep{Allison:2011}, and dynamically processing binaries
\citep{Parker:2011c}.  However, it
is unclear whether lower-mass clusters can also undergo this cool collapse phase within
1\,Myr.

We show typical examples of the fractal morphologies for all of our simulated clusters. We
model the ONC with a fractal of radius $\sim$ 1\,pc, an initial virial ratio of $Q = 0.3$
and a fractal dimension $D = 2.0$. In Fig.~\ref{ONC_morph} we show the fractal model for the
ONC (simulation ID = 4 in Table~\ref{cluster_props}) before dynamical evolution (panel(c))
and after 1\,Myr (panel(d)).

A key point to note in Fig.~\ref{ONC_morph} is that the Plummer sphere and fractal initial
conditions look very different at 0~Myr, but by 1~Myr the fractal has relaxed and the two
clusters appear very similar (the initially fractal cluster is slightly more compact).
The ONC fractal cluster models assume that the cluster is sub-virial (cool), which causes the cluster
to undergo a collapse, forming a dense core (see \citet{Allison:2010}
for details). 

As the
central density of IC348 is $1115 \pm 140$\,stars\,pc$^3$, significantly higher than most
star forming regions \citep[e.g.][]{Bressert:2010}, we test whether this cluster also
underwent a cool collapse phase. We also model IC348  with a fractal of radius $\sim$ 1\,pc,
an initial virial ratio of $Q = 0.3$ and a fractal dimension $D = 2.0$ (simulation ID = 10
in Table~\ref{cluster_props}), and typical morphologies are shown in Fig.~\ref{IC348_morph}.
The cluster before dynamical evolution is shown in Fig.~\ref{IC348-morph-c} and at 1\,Myr in
Fig.~\ref{IC348-morph-d}, with the centroid of the cluster shown by the red cross.

In contrast to IC348, Ophiuchus is relatively sparse, with  a central density of $610 \pm
180$\,stars\,pc$^3$. We elect to model this cluster as a fractal in virial equilibrium ($Q =
0.5$); even if this cluster is undergoing cool collapse, it will not reach its densest phase
until long after 1\,Myr. However, the map of the cluster (Fig.~\ref{fig:Oph_map}) shows a
moderate level of substructure in this cluster. We therefore set up the cluster with fractal
dimension $D = 2.0$ and a radius of $\sim$ 1\,pc (simulation ID = 14 in
Table~\ref{cluster_props}), and with typical morphologies as shown in Fig.~\ref{Oph_morph}.
The cluster before dynamical evolution is shown in Fig.~\ref{Oph-morph-c} and at 1\,Myr in
Fig.~\ref{Oph-morph-d}, with the centroid of the cluster shown by a red cross.

The Cham~I cluster appears to be highly sub-structured and relatively sparse (recall
Fig.~\ref{fig:Cha_map}). If clusters form with large amounts of sub-structure, which is then
subsequently erased by dynamics, then we can already hypothesise that this cluster has not
undergone much dynamical evolution. We model this cluster as a highly sub-structured fractal
($D = 1.6$) with radius 3\,pc, in virial equilibrium $Q = 0.5$ (simulation ID = 16 in
Table~\ref{cluster_props}),  and with typical morphologies as shown in Fig.~\ref{Cha_morph}.
The cluster before dynamical evolution is shown in Fig.~\ref{Cha-morph-a} and at 1\,Myr in
Fig.~\ref{Cha-morph-b}. As there is no well-defined centre of the cluster, we mark the
location in the cluster with the highest stellar surface density with a red circle.

Similarly, Taurus is also highly sub-structured, without a well defined centre. We  model
this cluster as a highly sub-structured fractal ($D = 1.6$) with radius 10\,pc, in virial
equilibrium $Q = 0.5$ (simulation ID = 18 in Table~\ref{cluster_props}),  and with typical
morphologies as shown in Fig.~\ref{Tau_morph}. The cluster before dynamical evolution is
shown in Fig.~\ref{Tau-morph-a} and at 1\,Myr in Fig.~\ref{Tau-morph-b}. Again, as there is
no well-defined centre of the cluster, we mark the location in the cluster with the highest
stellar surface density by a red circle.

\begin{table*}
\caption[bf]{Properties of our simulated clusters. From left to right,
  the cluster name, number of stars, morphology,
  fractal dimension ($D$) when relevant, the half-mass radius ($r_{1/2}$) 
  or fractal radius ($r_{\rm F}$), initial virial ratio $Q$, initial binary 
  fraction, and the symbol for the simulation in Figs.~\ref{ONC_sim}, \ref{IC348_sim}, 
\ref{Oph_sim}, \ref{Cha_sim}, \ref{Tau_sim} and \ref{all_sim}.}
\begin{center}
\begin{tabular}{|c|c|c|c|c|c|c|c|}
\hline 
Cluster & $N_{\rm stars}$ & Morphology & $D$ & $r_{1/2}$ or $r_{\rm F}$ & $Q$ & $f_{\rm bin}$ & symbol \\
\hline
ONC & 1500 & Plummer & -  & 0.1\,pc & 0.5 & 100\,per cent & $\times$ \\
ONC & 1500 & Plummer & -  & 0.4\,pc & 0.5 & field-like & $\lozenge$ \\
ONC & 1500 & Plummer & -  & 0.8\,pc & 0.5 & field-like & $\filledstar$ \\
ONC & 1500 & Fractal & 2.0  & 1\,pc & 0.3 & 100\,per cent & $\filledtriangleup$ \\
ONC & 1500 & Fractal & 2.0  & 1\,pc & 0.3 & 73\,per cent & $\bullet$ \\
ONC & 1500 & Fractal & 2.0  & 1\,pc & 0.3 & field-like & $\filledsquare$ \\
\hline 
IC348 & 260 & Plummer & -  & 0.1\,pc & 0.5 & 100\,per cent & $\times$ \\
IC348 & 260 & Plummer & -  & 0.4\,pc & 0.5 & field-like & $\lozenge$ \\
IC348 & 260 & Plummer & -    & 0.8\,pc & 0.5 & field-like & $\filledstar$ \\
IC348 & 260 & Fractal & 2.0  & 1\,pc & 0.3 & 100\,per cent & $\filledtriangleup$ \\
IC348 & 260 & Fractal & 2.0  & 1\,pc & 0.3 & 73\,per cent & $\bullet$ \\
IC348 & 260 & Fractal & 2.0  & 1\,pc & 0.3 & field-like & $\filledsquare$ \\
\hline
Oph & 300 & Plummer & -  & 0.1\,pc & 0.5 & 100\,per cent & $\times$ \\
Oph & 300 & Plummer & -  & 0.4\,pc & 0.5 & field-like & $\lozenge$ \\
Oph & 300 & Plummer & -  & 0.8\,pc & 0.5 & field-like & $\filledstar$ \\
Oph & 300 & Fractal & 2.0  & 1\,pc & 0.5 & 100\,per cent & $\smalltriangleup$ \\
Oph & 300 & Fractal & 2.0  & 1\,pc & 0.5 & 73\,per cent & $\circ$ \\
Oph & 300 & Fractal & 2.0  & 1\,pc & 0.5 & field-like & $\smallsquare$ \\
\hline
Cham~I & 200 & Fractal & 1.6  & 3\,pc & 0.5 & 100\,per cent & $\smalltriangleup$ \\
Cham~I & 200 & Fractal & 1.6  & 3\,pc & 0.5 & 73\,per cent & $\circ$ \\
Cham~I & 200 & Fractal & 1.6  & 3\,pc & 0.5 & field-like & $\smallsquare$ \\
\hline
Taurus & 300 & Fractal & 1.6  & 10\,pc & 0.5 & 100\,per cent & $\smalltriangleup$ \\
Taurus & 300 & Fractal & 1.6  & 10\,pc & 0.5 & 73\,per cent & $\circ$ \\
Taurus & 300 & Fractal & 1.6  & 10\,pc & 0.5 & field-like & $\smallsquare$ \\
\hline
\end{tabular}
\end{center}
\label{cluster_props}
\end{table*}

\subsection{Results}

Our ensembles of simulations with the symbols used in Figs.~\ref{ONC_sim}--\ref{all_sim} are
summarised in Table.~\ref{cluster_props}. In the figures we show the multiplicity after
1~Myr in the applicable separation ranges for each ensemble of simulations against the
stellar number density and the observed values for each cluster as the red datapoint.

In this subsection we will briefly review the results before discussing their implications
later.

\subsubsection{The ONC}

\begin{figure}
 \begin{center}
%\rotatebox{270}
{\includegraphics[scale=0.37]{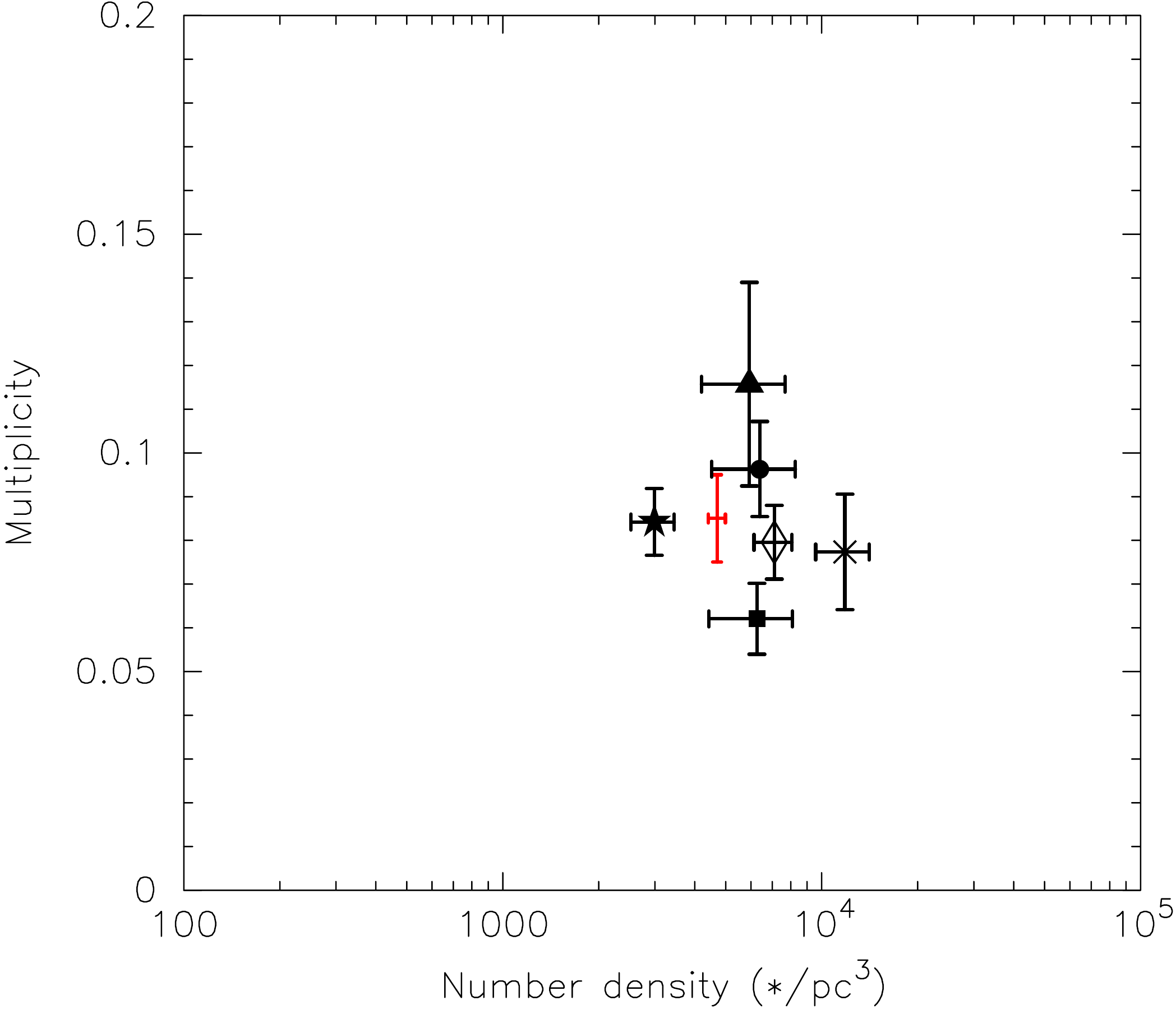}}
\caption[bf]{Multiplicity versus stellar density in the separation range 62 -- 620\,au for various simulated ONC-like clusters (see Table~\ref{cluster_props} for details). The 
observed ONC value is shown by the red datapoint.}
\label{ONC_sim}
\end{center}
\end{figure}

In Fig.~\ref{ONC_sim} we plot the binarity in the range 62 -- 620\,au as a function of
density within 0.25\,pc of the centroid of the cluster.  As can be
seen, none of the 
simulations are perfect fits, but none are terribly bad fits
either. The 73 per cent binary fraction collapsing fractal 
collapsing fraction just matches the observations.

\subsubsection{IC348}

\begin{figure}
 \begin{center}
%\rotatebox{270}
{\includegraphics[scale=0.37]{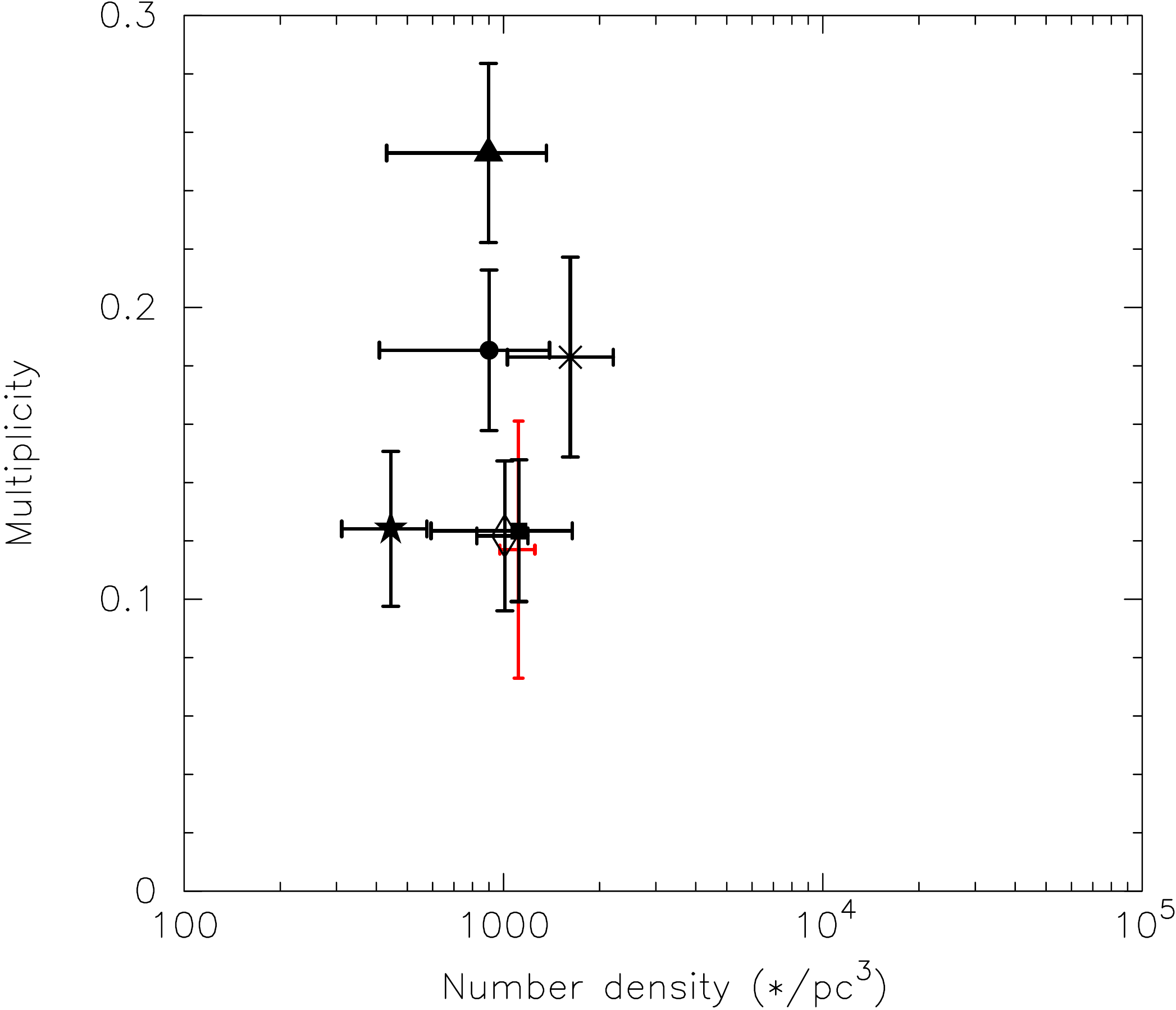}}
\caption[bf]{Multiplicity versus stellar density in the separation range 32 -- 830\,au for various simulated IC348-like clusters (see Table~\ref{cluster_props} for details). The 
observed value for IC348 is shown by the red datapoint.}
\label{IC348_sim}
\end{center}
\end{figure}

In Fig.~\ref{IC348_sim} we plot the binarity in the range 32 -- 830~au as a function of
density within 0.25\,pc of the centroid of the cluster.  For IC348 very good fits to the
observations are found for (a) a field-like initial binary fraction in a 0.4~pc radius
Plummer sphere, or (b) a collapsing fractal with a field-like binary
fraction.  Simulations with an initially 100 per cent binary
fraction are particularly poor fits.  The 73 per cent binary fraction
collapsing fractal simulation is, again, just consistent with the observations.

\subsubsection{Ophiuchus}

\begin{figure}
 \begin{center}
%\rotatebox{270}
{\includegraphics[scale=0.37]{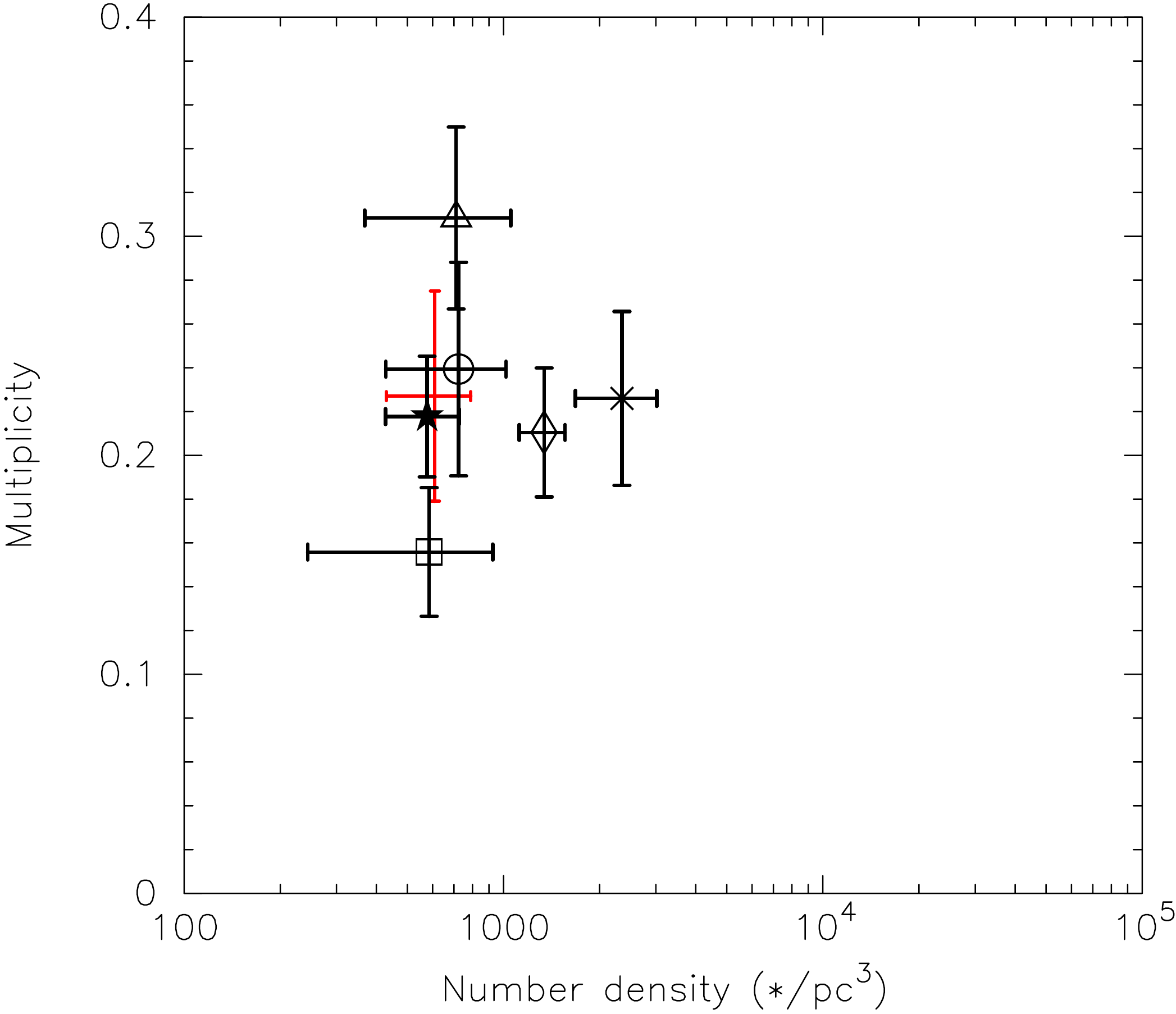}}
\caption[bf]{Multiplicity versus stellar density in the separation range 18 -- 830\,au for various simulated Ophiuchus-like clusters (see Table~\ref{cluster_props} for details). The 
observed value in Ophiuchus is shown by the red datapoint.}
\label{Oph_sim}
\end{center}
\end{figure}

In Fig.~\ref{Oph_sim} we plot the binarity in the range 18 -- 830~au as a function of
density within 0.25\,pc of the centroid of the cluster.  In Ophiuchus, a field-like binary
fraction, 0.8~pc radius Plummer sphere and a $Q=0.5$ fractal with a
73 per cent binary fraction are both reasonable fits. 

\subsubsection{Cham~I}

\begin{figure}
 \begin{center}
%\rotatebox{270}
{\includegraphics[scale=0.37]{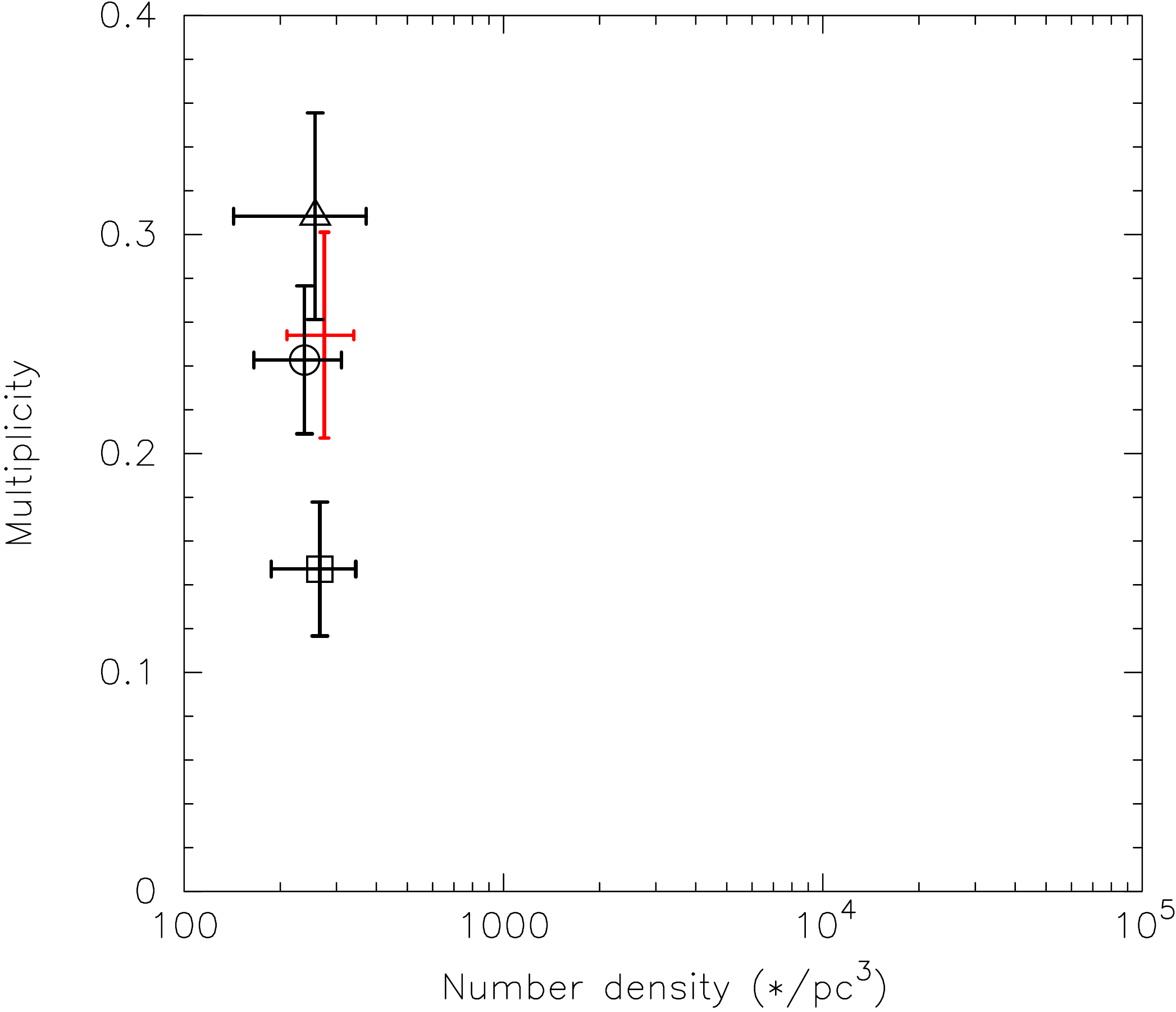}}
\caption[bf]{Multiplicity versus stellar density in the separation range 18 -- 830\,au for various simulated Chamaeleon-like clusters (see Table~\ref{cluster_props} for details). The 
observed value in Chamaeleon~I is shown by the red datapoint.}
\label{Cha_sim}
\end{center}
\end{figure}

Fig.~\ref{Cha_sim} shows the binarity in the range 18 -- 830~au as a function of density
within 0.25\,pc of the star with the highest stellar surface density in the cluster.  As
Cham~I is rather clumpy we only model it as a fractal.  Neither the 100 per cent or the 
field-like binary fractions are particularly good fits, but the 73 per
cent $Q=0.5$ fractal is a rather good fit.

\subsubsection{Taurus}

\begin{figure}
 \begin{center}
%\rotatebox{270}
{\includegraphics[scale=0.37]{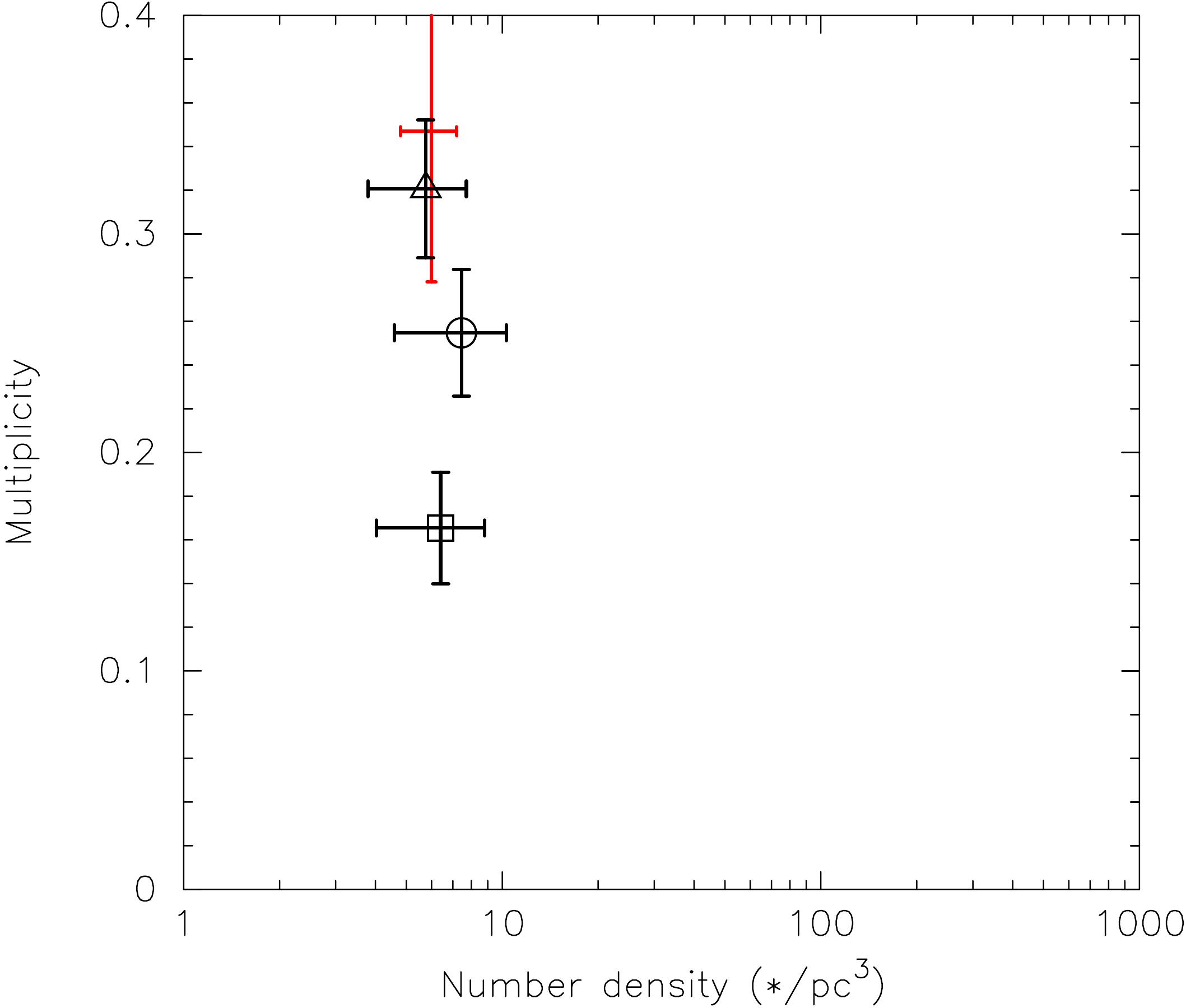}}
\caption[bf]{Multiplicity versus stellar density in the separation range 18 -- 830\,au for the simulated Taurus-like clusters (see Table~\ref{cluster_props} for details). The 
observed value in Taurus is shown by the red datapoint.}
\label{Tau_sim}
\end{center}
\end{figure}

Finally, in Fig.~\ref{Tau_sim} we show the binarity in the range 18 -- 830~au as a function
of density within 1~pc of the star with the highest stellar surface density in the cluster. 
As with Cham~I we only model Taurus as a fractal.  In this case, the fractal with 100 per
cent initial binary fraction is by far the best fit, but the 73 per
cent binary fraction, $Q=0.5$ fractal is also consistent with the
observations.

%%%%%%%%%%%%%%%%%%%%%%%%%%%%%%%%%%%%%%%%%%%%%%%%%%%%%%%%%%%%%%%
\section{Discussion}

In this paper we wish to address the question of the universality of star formation.  In
particular we wish to examine if the morphologies, densities and binary fractions (within a
given separation range) of the observed clusters can be matched by $N$-body simulations with
statistically the same initial conditions (with only the number of stars varying).

We have constructed directly comparable observational samples of the binary fractions of
Taurus, Cham~I, Ophiuchus, IC348, and the ONC.  In as far as is possible these samples cover
the same physical separation ranges for stars in the same mass range, and with the same
companion magnitude differences.  Such samples are crucial to ensure we are comparing
like-with-like.

In our simulations we have aimed to start with a wide variety of initial conditions, but to
try to match the observed morphologies, densities and binary fractions of the observed
clusters at an age of 1~Myr.  

\subsection{A single model?}

\begin{figure}
 \begin{center}
{\includegraphics[scale=0.37]{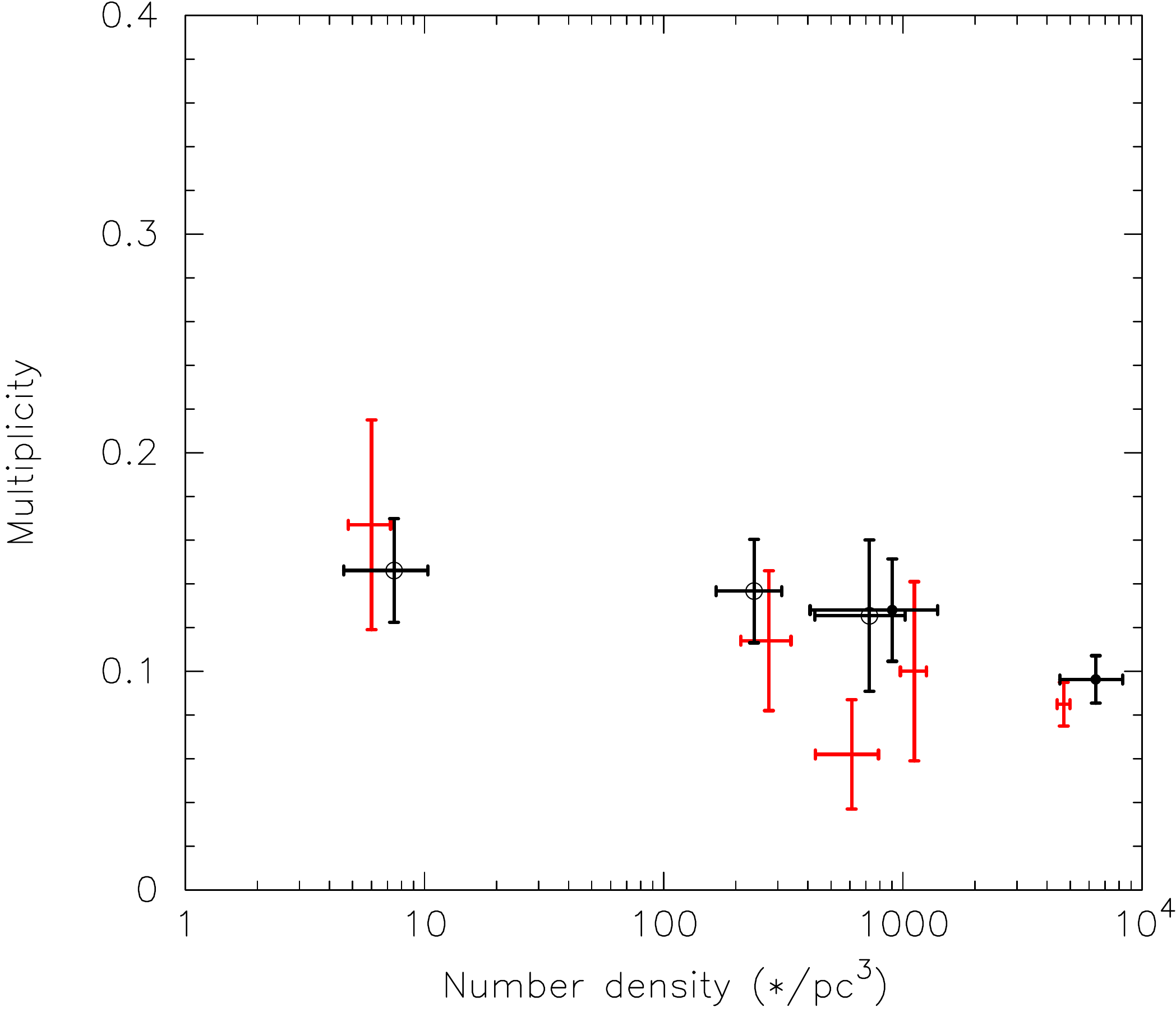}}
\caption[bf]{Multiplicity versus stellar density in the separation range 62 -- 620\,au 
for all five clusters with a total binary fraction of 73 per cent. The observed values are 
shown by the red datapoints, the sub-virial clusters shown by filled circles and those in 
virial equilibrium shown by open circles.}
\label{all_sim}
\end{center}
\end{figure}

Is there a single model that reproduces all of the observations?  The answer is almost, but
with several caveats.

The {\em best} fits for the clusters are all different.  The best fits for the ONC and
Ophiuchus suggest a fairly large Plummer sphere with an initially field-like binary fraction
as the initial conditions.  For IC348 a field-like binary fraction is certainly preferred,
but with no preference between Plummer or fractal initial conditions.  The current state of
Taurus and Cham~I are clearly clumpy (modelled as fractals, although they may well not be
truly fractal) with a clear preference for a 100 per cent binary fraction in Taurus, and a
somewhat lower fraction in Cham~I.

The first thing that we can say is that {\em a universal model must start with
substructure}. Taurus and Cham~I are clearly substructured, and so any universal model must
also be substructured.  However, {\em a universal model must rapidly erase its substructure}
in order to match the fairly uniform appearance of Ophiuchus, IC348, and the ONC.

With numerical experiments we are able to construct a single model that (just) fits all of
the clusters (see Fig.\,\ref{all_sim}).  This model is fractal with a 73 per cent initial {\em
total} binary fraction. This 73 per cent binary fraction is the initial binary fraction
across all separation ranges. Note that we assume that the initial separation distribution
is a log normal with $\mu_{{\rm log}~a} = 1.53$ and $\sigma_{{\rm log}~a} = 1.57$ (\idest\
Duquennoy \& Mayor 1991 for field G-dwarfs).  We will return to a discussion of this
assumption later.

However, in order to reproduce the observations we require two different initial virial
ratios, $Q=0.3$ for Ophiuchus, IC348, and the ONC, and $Q=0.5$ for Taurus and Cham~I.  The
reason for this is that we need to erase substructure in Ophiuchus, IC348, and the ONC (so
needing a collapse), but retain substructure in Taurus and Cham~I.

This might not be a great problem.  We have assumed an age of 1\,Myr for all clusters, but
this is an approximation.  The ONC probably has an age of 2--3\,Myr
\citep{Mayne:2008,DaRio:2010} and so has had longer to erase its substructure.  However,
sub-virial clumpy initial conditions are required if we wish to reproduce features of the
ONC such as the mass segregation \citep[see][]{Allison:2009b}.

The fractal dimensions used are also slightly different, $D=2$ for Ophiuchus, IC348, and the
ONC, and $D=1.6$ for Taurus and Cham~I. Higher fractal dimensions begin smoother and so can
erase their substructure more easily.  Again, we feel that this may not be too great an
obstacle to overcome as we did not test many different fractal dimensions.  Largely this is
because we do not believe the initial distributions to be {\em truly} fractal, rather the
fractal is a useful numerical tool when constructing substructured distributions.

Therefore it might be possible to construct a universal model with high clumpiness (low
fractal dimension) close to, but below, virial equilibrium which allows the substructure to
be erased rapidly.  To be consistent with the clusters discussed here, the initial binary 
fraction must be high, but not 100 per cent.

Taurus and Cham~I present something of a problem for a universal model.  Both Taurus and
Cham~I are substructured and must be dynamically fairly unevolved.  However, Cham~I has a
binary fraction much closer to the presumably dynamically processed Ophiuchus.  Is Cham~I
much more dynamically evolved than Taurus (similar to Ophiuchus), but for some reason has
not yet erased its substructure?

\subsection{Assumptions about the binary separation distribution}

It must be remembered that we are only constrained by observations in the separation
range(s) which are observed.  For all five clusters we can only compare in the range 62 --
620~au.  Rather frustratingly, this range is dominated by `intermediate' binaries, that is
binaries that are neither `hard' nor `soft' and whose survival depends on the exact details
of the dynamical evolution and if they were `unlucky' enough to have had a destructive
encounter or not.  Soft binaries are destroyed within a crossing time \citep[although they 
may appear and disappear in clusters as they can be re-formed,][]{Moeckel:2011}.  Hard
binaries are almost never processed.

Whether an intermediate binary is destroyed, firstly depends on the velocity dispersion of
the cluster (which sets the separation range which is susceptible to destruction).  It also
depends on the encounter probability (itself a function of the velocity dispersion and
stellar density) which sets the chance of a destructive encounter occurring. Indeed, in a low
enough density environment (such as the field) formally soft binaries can survive as
encounters are extremely rare.

Therefore, even though we find that a 73 per cent initial binary fraction is the best fit,
this assumes a G-dwarf field-like log-normal distribution across the whole separation range.
Actually the observations tell us nothing about binaries with separations $<62$~au or
$>620$~au in all clusters.  Thus our 73 per cent total initial binary fraction corresponds
to a $\sim 17$ per cent initial binary fraction in the range 62--620~au, or $\sim 30$ per
cent in the range 18--830~au.  However, outside, and especially below, our observed
separation ranges we have no information and the 73 per cent value should be taken with
caution.  As an extreme example, the ONC {\em could} have a 100 per cent binary fraction if
it had 90 per cent of its stars in 10~au binaries.  Observationally, we cannot refute this
claim.

%%%%%%%%%%%%%%%%%%%%%%%%%%%%%%%%%%%%%%%%%%%%%%%%%%%%%%%%%%%%%%%%%%
%%%%%%%%%%%%%%%%%%%%%%%%%%%%%%%%%%%%%%%%%%%%%%%%%%%%%%%%%%%%%%%%%%

%%%%%%%%%%%%%%%%%%%%%%%%%%%%%%%%%%%%%%%%%%%%%%%%%%%%%%%%%%%%%%%%%%%%%%%%
\section{Conclusions}

We have produced comparable selections of binaries in five clusters: Taurus, Ophiuchus,
Cham~I, IC348, and the ONC for stars between 0.1 and 3 $M_\odot$.  The multiplicity in each
cluster has been determined to the same mass limits, mass ratio sensitivity, and within the
same separation ranges allowing them to be compared directly with one another.  For all five
clusters we have multiplicities in the range 62 -- 620~au, for all but the ONC we have a
range of 32 -- 830~au, and for Taurus, Ophiuchus, and Cham~I we have a range of 18 --
830~au.

We find in common with previous work that in the range 62 -- 620~au Taurus has a significant
excess of binaries compared to the ONC ($16.7 \pm 4.8$ per cent versus $8.5 \pm 1.0$ per
cent).  We find that the trend of decreasing binary fraction with increasing density is
driven solely by the high binary fraction of Taurus.  Ophiuchus, Cham~I, IC348, and the ONC
cover a factor of 17 in density but show no discernable trend.  However, we note that a
global `density' is a rather difficult term to define for sub-structured clusters such as
Taurus and Cham~I.

We perform a large ensemble of $N$-body simulations of Plummer spheres and fractal
distributions with different initial (log-normal) binary fractions.  We then `observe' the
simulations in the same separation ranges for the same masses and with the same mass ratio
sensitivity as the real observations.

We are able to find one set of simulations that roughly reproduce all of the observations: a
fractal with a total binary fraction of 73 per cent (albeit with slightly different fractal
dimensions and virial ratios).  Universal initial conditions must therefore be clumpy as
both Taurus and Cham~I are clearly not smooth.  Universal initial conditions must also have
a higher binary fraction in the range 62--620\,au than currently observed (in all but
Taurus) as dense clusters must process some of their binary population in this range. To test 
this scenario, binary surveys probing closer separations ($\sim$10--60\,au) are required.

\section*{Acknowledgments} 

We thank the anonymous referee for a prompt and helpful report. We acknowledge support
for this work from European Commission Marie Curie Research Training Network CONSTELLATION
(MRTN-CT-2006-035890) and the Exeter Astrophysics visitor grant. R.R.K. is supported by a
Leverhulme research project grant (F/00 144/BJ).  The $N$-body simulations in this work were
performed on the \texttt{BRUTUS} computing cluster at ETH Z{\"u}rich; and the
\texttt{Iceberg} computing cluster, part of the White Rose Grid computing facilities at the
University of Sheffield.

\bibliographystyle{mn2e}
\setlength{\bibhang}{2.0em}
\setlength\labelwidth{0.0em}
\bibliography{bib}

\bsp
\label{lastpage}

\end{document}